\documentclass[sigconf,10pt]{acmart}
\usepackage[english]{babel}
\usepackage{blindtext}
\usepackage{comment}
\usepackage{dsfont}
\usepackage{CJKutf8}
\usepackage{mathrsfs,amsmath}
\usepackage{mathtools}
\usepackage{aligned-overset}

\usepackage{kantlipsum}
\usepackage{multirow}
\usepackage{subfigure}
\usepackage{colortbl}
\usepackage{listings}

\usepackage{blindtext}

\usepackage{pifont}% http://ctan.org/pkg/pifont

\usepackage{tikz}
\newcommand*\circled[1]{\tikz[baseline=(char.base)]{
            \node[shape=circle,draw,inner sep=0.7pt] (char) {\textbf{\textcolor{black}{#1}}};}}
            
\usepackage{array}
\newcolumntype{C}[1]{>{\centering\arraybackslash}p{#1}}

\renewcommand\footnotetextcopyrightpermission[1]{} % removes footnote with conference info
\setcopyright{none}
\acmDOI{}

\acmISBN{}

\acmPrice{}

\newcommand{\ours}{{m4}\xspace}

\newcommand{\nsthree}{{ns-3}\xspace}
\newcommand{\flowsim}{{flowSim}\xspace}
\newcommand{\cache}{{CacheFollower}\xspace}
\newcommand{\hadoop}{{Hadoop}\xspace}
\newcommand{\web}{{WebServer}\xspace}
\newcommand{\dctcp}{{DCTCP}\xspace}
\newcommand{\timely}{{TIMELY}\xspace}
\newcommand{\dcqcn}{{DCQCN}\xspace}
\newcommand{\errortail}{{p90}\xspace}
\newcommand{\errormeanreduce}{{45.3\%}\xspace}
\newcommand{\errortailreduce}{{53.0\%}\xspace}
\newcommand{\errormeanours}{{7.5\%}\xspace}
\newcommand{\errortailours}{{19.5\%}\xspace}
\newcommand{\errortailsldnours}{{13.3\%}\xspace}
\newcommand{\errormeanflowsim}{{13.7\%}\xspace}
\newcommand{\errortailflowsim}{{41.5\%}\xspace}
\newcommand{\errortailsldnflowsim}{{23.2\%}\xspace}
\newcommand{\speedupovernsthree}{{104$\times$}\xspace}
        
\newcolumntype{"}{!{\vrule width 2pt}}

\definecolor{codeblue}{rgb}{0,0,1}
\usepackage[linesnumbered,ruled,noend]{algorithm2e}
\let\oldnl\nl% Store \nl in \oldnl
\newcommand{\nonl}{\renewcommand{\nl}{\let\nl\oldnl}}% Remove line number for one line
\SetCommentSty{mycommfont}
\SetKwInput{KwInput}{Input}                % Set the Input
\SetKwInput{KwOutput}{Output}              % set the Output

\def\compactify{\itemsep=0pt \topsep=0pt \partopsep=0pt \parsep=0pt}
\let\latexusecounter=\usecounter
\if 0

\fi

\usepackage{enumitem}
\setlist{nolistsep}
\usepackage{cleveref}

\begin{document}
\title{\ours: A Learned Flow-level Network Simulator}

\author{\large Chenning Li$^{1}$, Anton A. Zabreyko$^{1}$, Arash Nasr-Esfahany$^{1}$, Kevin Zhao$^{2}$, \\ Prateesh Goyal$^{3}$, Mohammad Alizadeh$^{1}$, and Thomas Anderson$^{2}$\\{\vspace{6mm}\large $^{1}$MIT CSAIL, $^{2}$University of Washington, $^{3}$Microsoft Research}}\vspace{2mm}

\renewcommand{\shortauthors}{X.et al.}

\begin{abstract}
Flow-level simulation is widely used to model large-scale data center networks due to its scalability. 
Unlike packet-level simulators that model individual packets, flow-level simulators abstract traffic as continuous flows with dynamically assigned transmission rates. 
While this abstraction enables orders-of-magnitude speedup, it is inaccurate by omitting critical packet-level effects such as queuing, congestion control, and retransmissions.

We present \ours, an accurate and scalable flow-level simulator that uses machine learning to learn the dynamics of the network of interest. 
At the core of \ours lies a novel ML architecture that decomposes state transition computations into distinct spatial and temporal components, each represented by a suitable neural network. 
To efficiently learn the underlying flow-level dynamics, \ours adds dense supervision signals by predicting intermediate network metrics such as remaining flow size and queue length during training.
\ours achieves a speedup of up to \speedupovernsthree over packet-level simulation.
Relative to a traditional flow-level simulation, \ours reduces per-flow estimation errors by \errormeanreduce (mean) and \errortailreduce (\errortail). 
For closed-loop applications, \ours accurately predicts network throughput under various congestion control schemes and workloads.

\end{abstract}

\maketitle
\pagestyle{plain}

\section{Introduction}\label{sec-introduction}
Packet-level simulators~\cite{ns3,omnet,htsim} are popular in networking research, but face significant scalability challenges for modeling large-scale data center networks. 
Recent work improves the scalability of packet-level simulation using machine learning~\cite{mimicnet, deepqueuenet, li2024m3}, approximation techniques~\cite{zhao2023scalable}, and better parallelization~\cite{dons}. 
However, these approaches have some key limitations. 
Most methods continue to operate at the packet level~\cite{mimicnet, deepqueuenet, zhao2023scalable, dons}, which inevitably becomes harder to scale as network speeds increase~\cite{li2024m3}. 
Others model aggregate performance at the distributional level but not the behavior of individual flows~\cite{mimicnet, li2024m3, zhao2023scalable}, limiting their applicability in tasks that require more fine-grained information, such as modeling distributed applications.

Flow-level simulation has emerged as a scalable alternative to packet-level simulation for modeling data center networks. 
Unlike packet-level simulators, flow-level simulators abstract traffic into continuous data streams with time-varying rates. 
By dynamically assigning flow rates based on bandwidth-sharing policies (e.g., max-min fairness), they enable fast analysis while capturing the first-order impact of bandwidth contention and congestion control. 
Moreover, flow-level simulators provide a natural abstraction for modeling communication in distributed applications (\S\ref{subsec:motivation-1}). 
For these reasons, many research and industrial systems use flow-level models to characterize network performance, e.g., in traffic engineering~\cite{hong2013achieving,jain2013b4,krishnaswamy2023onewan}, distributed machine learning (ML) training, and inference~\cite{narayanan2020heterogeneity,won2023astra,rashidi2020astra,cassini,cao2024crux,sharma2024gpu,cho2024llmservingsim}.

Despite its advantages, flow-level simulation is inherently inaccurate because it abstracts or ignores key packet-level details, such as congestion control, queuing, packet loss, and retransmissions~\cite{zhao2023scalable,li2024m3}.
For example, flow-level simulators typically do not model queues, causing them to systematically underestimate flow completion times (FCTs), especially for short flows and at the tail (\S\ref{subsec:motivation-1}, \Cref{tab:preliminary}). 
\textit{Our goal is to develop a model that combines the speed of flow-level simulation with accuracy comparable to packet-level models.}

This paper proposes \ours, an accurate and scalable flow-level simulator that uses ML to model the behavior of individual network flows (termed ``flow-level dynamics'') as they traverse a network topology. 
\ours is trained using ground-truth data generated by a packet-level simulator such as \nsthree, though our techniques can in theory be used to learn flow-level behavior in a real network.  
Like existing flow-level models, \ours provides a simple interface where applications specify flow characteristics (e.g., source, destination, network path, and size) and receive completion notifications. 
Unlike traditional approaches, \ours learns the flow-level dynamics directly from data, capturing the effect of complex factors such as congestion control behavior, queuing delays, and packet retransmissions.

To understand \ours, consider the problem of estimating FCTs for an arbitrary sequence of flows. 
This can be framed as a \textit{sequence-to-sequence} modeling problem, where the input is a sequence of flow arrival events $(f_1, f_2, ..., f_n)$ denoting the start time, size, and network path for each flow, and the output is the sequence of FCTs $(FCT_1, FCT_2, ..., FCT_n)$.
It is tempting to apply standard sequence models like Long Short-Term Memory (LSTM)~\cite{lstm} or Transformer~\cite{transformer} to this problem.
However, sequence models have two fundamental limitations: 1) They use a fixed-sized hidden state vector to represent the global state of the network, making it difficult to scale the topology and number of flows.
2). They are poorly suited to processing topology or routing information, which are inherently graph-structured data. 

To address these challenges, \ours employs a novel neural network architecture inspired by the computational structure of traditional flow-level simulators. 
\ours maintains a ``hidden state'' for each flow and link, with the set of hidden states of all active flows and links representing the global state of the network at a given time. 
\ours then learns a state transition function that updates the state across flow-level events (arrivals and departures) and output functions that generate observables such as FCTs and queue lengths based on the state. 
A key idea, based on the structure of existing flow-level simulators, is to decompose the complex state transition function into two components, each represented by a neural network architecture suitable for that component: 
1) a {\em spatial model}, capturing the interactions between flows and network links at each time step using a Graph Neural Network (GNN)~\cite{hamilton2017inductive}. 
2) a {\em temporal model}, capturing how each hidden state evolves over time using a sequence model (see~\S\ref{subsec:motivation-3} and \S\ref{sec-system-design} for details).

A major challenge in training a complex model like \ours is to generalize effectively across diverse network scenarios.
This requires learning the true underlying rules governing flow-level dynamics rather than spurious correlations that happen to fit the training data~\cite{dietmuller2022new}. 
Although the primary goal of \ours is to predict FCT information, relying solely on per-flow supervision by minimizing the loss between predicted and actual FCTs provides a sparse and inadequate learning signal.
Large flows, in particular, may encounter hundreds or thousands of flow-level events before completion, with each event impacting the state of the network and the behavior of other flows.
Learning this process from FCTs alone—a single scalar per flow—is extremely challenging. 

To learn generalizable flow-level dynamics from data, \ours's second idea is to add dense supervision signals at each flow-level event during training.
The rationale is that \ours maintains hidden states for flows and links which are meant to encompass all relevant information about their behavior at each point in time. 
At each flow-level event, \ours queries these hidden states to predict intermediate metrics, such as the remaining flow size of each active flow or the queue occupancy of relevant links.
To train its ML models, \ours minimizes a combined loss function that includes FCT, remaining size, and queue occupancy prediction losses.
By providing per-event supervision rather than relying solely on per-flow FCT predictions, \ours achieves a more detailed understanding of flow-level dynamics and generalizes more effectively across diverse network scenarios.

We train \ours on diverse synthetically generated simulation scenarios using small 32-host fat-tree topologies.
These synthetic scenarios span complex flow-level dynamics, including flow size variations, burstiness levels, congestion control protocols, network loads, and buffer sizes.
We validate \ours against production workloads running on network topologies with up to 6144 hosts, modeled after Meta's data center network~\cite{meta_fabric}. 
We evaluate the estimation error in predicting the FCT slowdown for each individual flow compared to ns-3, where FCT slowdown is defined as the ratio of the flow completion time to the ideal completion time on an unloaded network.
We summarize our evaluation results below.
\begin{itemize}[leftmargin=*]
\item On a 384-rack, 6144-host fat-tree topology, \ours reduces the mean (\errortail) estimation error compared to a standard flow-level simulator (\flowsim) by 47\% (69\%) to 77\% (81\%) across different scenarios.
\ours also delivers a speedup of up to \speedupovernsthree compared to \nsthree.
\item Using a 64-rack, 1024-host fat-tree topology with diverse production workloads and network configurations, \ours reduces the mean (\errortail) error for per-flow FCT slowdown estimation by \errormeanreduce (\errortailreduce) compared to \flowsim. 
\ours achieves mean (\errortail) errors of \errormeanours (\errortailours), significantly outperforming flowSim's error of \errormeanflowsim (\errortailflowsim).  
\item On a machine with a 64-core processor and an A100 GPU, \ours can simulate a 49,152-server fat-tree topology with 10Gbps links running for a second in 24 minutes (75.2$\times$ speedup compared to \flowsim).
Although this is the largest scale in our experiments, our results suggest that \ours's speedup compared to \flowsim increases super-linearly with network size.
\item For closed-loop traffic applications, \ours dynamically processes flows from the application, accurately forecasting individual flow completion times.
Compared to \flowsim, \ours improves throughput prediction, reducing the mean (\errortail) error from 28.1\% (49.9\%) to 11.5\% (22.3\%).
\end{itemize}

{\ours}'s code is available at \href{https://github.com/netiken/m4}{https://github.com/netiken/m4}.
This work does not raise any ethical issues.

\section{Background and Motivation}
\label{sec-background-motivation}

\begin{table*}%[ht]
 \small
  \centering
  \begin{tabular}{|c|c|c|c|c|c|c|>{\columncolor{yellow!20}}c|>{\columncolor{yellow!20}}c|>{\columncolor{yellow!20}}c|c|c|}
    \hline
    \multirow{2}{*}{\textbf{Scenario}} & \multirow{2}{*}{\textbf{Workload}} & \multirow{2}{*}{\textbf{Max load}} & \multirow{2}{*}{\textbf{Oversub}} & \multirow{2}{*}{\textbf{CCA}} & \multicolumn{3}{c|}{\textbf{Wallclock time}} & \multicolumn{2}{c|}{\textbf{Per-flow error}} &  \multicolumn{2}{c|}{\textbf{Tail slowdown}} \\
    \cline{6-12}
    & & & & &\nsthree & \flowsim & Speedup & Mean & \errortail &\nsthree & \flowsim   \\
    \hline
    {1} & \cache  & {33.45\%} & {4-to-1} & \dctcp & 30.6mins & 0.8s & 2295$\times$ & {13.09\%} & {33.76\%} &4.14 & 3.07   \\
    \cline{1-12}
    {2} & \hadoop  & {57.60\%} & {4-to-1} & \timely & 31.0mins & 1.1s & 1691$\times$ & {11.00\%} & {36.57\%} & 3.78 & 2.84\\
    \cline{1-12}
    {3} & \hadoop & {73.72\%} & {1-to-1} & \dctcp & 43.4mins & 3.5s & 745$\times$ & {20.52\%} & {57.84\%} & 7.81 & 3.92  \\
    \hline
  \end{tabular}
  \caption{\small Comparison of the per-flow flow completion time (FCT) slowdown (sldn) and wallclock runtimes for 20K flows for \nsthree and \flowsim. Tail slowdown refers to the estimated $99^{th}$ percentile FCT slowdown. Configuration is the same as Section~\ref{subsec-evaluation-2}.}
  \vspace{-4mm}
  \label{tab:preliminary}
\end{table*}

\subsection{Why flow-level simulation?}
\label{subsec:motivation-1}
Flow-level simulators have emerged as an alternative to packet-level simulators such as \nsthree~\cite{ns3}, OMNET++~\cite{omnet}, and htsim~\cite{htsim} for modeling data center networks. 
Modern data center networks have hundreds of thousands of servers, network links, and network switches. 
As link speeds continue to increase, simulating data center networks at scale at the packet level has become impractical.

Flow-level simulators model ``fluid'' flows sending at time-varying rates and assign flow rates upon flow arrivals and departures according to a bandwidth sharing policy such as max-min fairness. 
Flow-level models have gained popularity in the industry and research community for two key reasons.
First, flow-level simulation is much faster than packet-level simulation. 
By abstracting away packet-level details such as queuing, retransmissions, and protocol processing, flow-level simulators process significantly fewer events while still capturing the first-order behavior of bandwidth contention and congestion control. 
Thus, many systems use flow-level models to characterize network performance, e.g., in traffic engineering~\cite{hong2013achieving,jain2013b4,krishnaswamy2023onewan} and GPU scheduling~\cite{narayanan2020heterogeneity,won2023astra,rashidi2020astra,cassini,cao2024crux,sharma2024gpu,cho2024llmservingsim}. 
Second, flow-level models offer a suitable interface for modeling distributed applications.
Standard transport protocols and communication libraries (e.g., gRPC~\cite{grpc}) hide packet-level details; applications send/receive messages, observing message start/completion events but not the underlying packet-level events. 
Thus, flow-level simulators are a convenient, scalable way to model communication in distributed applications. 
For example, ASTRA-sim~\cite{rashidi2020astra,won2023astra}, a state-of-the-art ML training simulator, takes distributed ML job traces as input, extracts collective communication operations, and converts them into a sequence of network flows.
These flows are then sent to a flow-level simulator backend to model network interactions and estimate communication times. 

The improved speed of flow-level simulators comes at the expense of accuracy. 
To illustrate the inherent tradeoffs, we compare packet-level (\nsthree) and flow-level (\flowsim) models in three simulation scenarios (\Cref{tab:preliminary}). 
\flowsim~\cite{massoulie1999bandwidth,namyar2023solving} is a simple flow-level simulator that computes flow completion times (FCT) by assigning max-min fair rates to all active flows at each time. 
The scenarios simulate a 64-rack, 1024-host fat-tree topology with various production workloads from Meta~\cite{fb-network}, maximum link loads, oversubscription levels, and congestion control algorithms~\cite{dctcp,mittal2015timely}  (detailed setup in \S\ref{subsec-evaluation-1}).
For each scenario, we simulate 20,000 flows with Equal-Cost Multi-Path (ECMP) routing using both \nsthree and \flowsim.
\Cref{tab:preliminary} summarizes the wallclock runtime of \flowsim and \nsthree, and the accuracy of \flowsim's FCT estimates relative to \nsthree in each scenario. 
Throughout this paper, we normalize the FCT of each flow by the minimum possible FCT in an unloaded network, which we refer to as {\em FCT slowdown}.

The results highlight two key findings:
1) \flowsim delivers a speedup of 745$\times$ to 2295$\times$ over \nsthree, reducing simulation runtime from tens of minutes to just a few seconds.
2) However, \flowsim is inaccurate. 
We calculate the relative estimation error of \flowsim for FCT slowdown on a per-flow basis, compared to \nsthree.
The mean (\errortail) error ranges from 11.00\% (33.76\%) to 20.52\% (57.84\%).
Furthermore, \flowsim significantly underestimates the tail (99th-percentile) FCT slowdown of each scenario, with errors ranging from 24.86\% (3.78 vs. 2.84) to 49.88\% (7.81 vs. 3.92) compared to \nsthree. 
As observed in prior work~\cite{zhao2023scalable,li2024m3}, this is because \flowsim cannot capture the effect of queuing delays and transport protocol intricacies, especially for short flows. 

These findings motivate \ours, a learned network simulation model that combines the speed of flow-level models with the accuracy of packet-level simulations.

\subsection{Learning flow-level models from data}
\label{subsec:motivation-2}
Our goal is to learn a dynamical system model for flow-level simulation from data.
This data comprises sequences of flow-level observations, such as flow arrivals and departures, and measurements of the trajectory of various quantities (e.g., remaining flow sizes, queue lengths) over time. 
Our hypothesis is that a learned model can capture the effect of complex factors, such as congestion control dynamics, packet loss, and retransmissions on flow-level performance {\em without operating at the packet level}. 
To train a model that generalizes, we collect training data under diverse network topologies and workloads. 
Our current system instruments \nsthree to collect training data, but our methodology applies to learning a model of any target environment.

A general flow-level model takes as input a sequence of flow-level events \( I = \{E(t_1), E(t_2), ..., E(t_i)\} \), where each \( E(t_i) \) corresponds to a flow arrival or departure occurring at time $t_i$. 
As the simulation progresses, the model produces a sequences of network states \( X = \{X(t_1), X(t_2), ..., X(t_i)\} \) (which are internal to the model) and a sequence of network observables \( O = \{O(t_1), O(t_2), ..., O(t_i)\} \), defined as follows:
\begin{equation}
\begin{aligned}
    X(t_i) &= \big\{s_f \mid f \in \mathcal{F}(t_i)\big\} \cup \big\{p_q \mid q \in \mathcal{Q}(t_i)\big\}, \\
    O(t_i) &= \big\{o_f \mid f \in \mathcal{F}(t_i)\big\} \cup \big\{o_q \mid q \in \mathcal{Q}(t_i)\big\}, 
\end{aligned}
\label{eq:flow_level_model}
\end{equation}
where \( s_f \) denotes the state of each active flow \( f \) in the flow set \( \mathcal{F}(t_i) \), and \( p_q \) represents the state of each switch queue \( q \) in the queue set \( \mathcal{Q}(t_i) \). 
The observables \( o_f \) include flow-level metrics such as the remaining flow sizes, while \( o_q \) represents link-level metrics such as the queue lengths at time \( t_i \). 
The model defines the evolution of network states and observables using the following discrete-time dynamical system:
\begin{equation}
\begin{aligned}
    X(t_i) &= f(X(t_{i-1}), E(t_i), t_i - t_{i-1}),  \\
    O(t_i) &= g(X(t_i)).  
\end{aligned}
\label{eq:general_model}
\end{equation}
Here, $f()$ is a {\em state transition function}, and $g()$ is an {\em output function} that expresses the observables in terms of the state. 
Our problem statement is to learn this model (i.e., functions $f$ and $g$) from a diverse dataset of example input and output sequences $\mathcal{D} = \bigcup_i \{I_i \rightarrow O_i\}$. 

A standard approach to this task is to pose it as a {\em sequence-to-sequence (seq2seq)} modeling problem. 
Sequence models like Long Short-Term Memory (LSTM)~\cite{lstm} maintain a ``hidden state'',  represented as a fixed-length vector, to model \( X(t_i) \), and update it as they process the input sequence $I$ and predict the output sequence $O$, one element at a time. 
However, applying standard sequence models to our problem presents two key challenges:
\begin{enumerate}[leftmargin=*]
    \item \textbf{Fixed-length representation constraint:} Sequence models use a fixed-size hidden state vector to represent the global state of the system, making it infeasible to represent network scenarios across different scales (e.g., tens to thousands of hosts and active flows) using the same model.  
    \item \textbf{Encoding topology and routing:} The network topology and routing information (i.e., the path taken by each flow) is inherently graph-structured information, and it is unclear how to encode it within a sequence model. 
    This is particularly challenging because we want the model to generalize across different network topologies. 
    Thus, a ``flat'' representation of the topology and routes using adjacency matrices or lists will not work. 
    
\end{enumerate}

\begin{figure}%[!t]
\centering
\includegraphics[width=0.99\columnwidth]{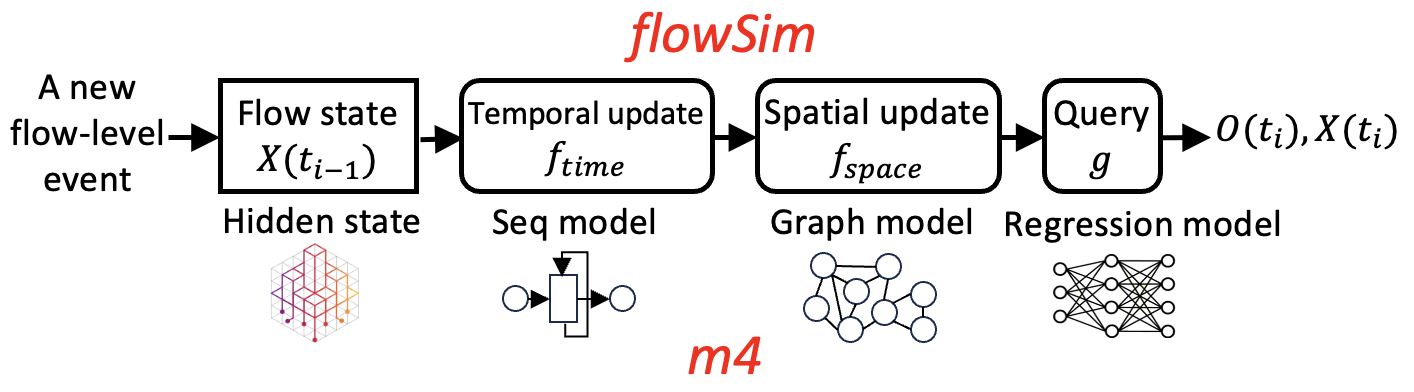}\label{subfig:motivation-2}
\caption{\small \ours mimics the computational structure of \flowsim but replaces its components with learnable modules.}  
\label{fig:motivation}
\vspace{-4mm}
\end{figure}

\subsection{Key Insight}
\label{subsec:motivation-3}
The crux of what make it difficult to define a parameterized architecture for learning model~\eqref{eq:general_model} is that the state transition function $f$ combines two forms of computation: 
1) {\em spatial modeling}, which considers the interactions of flows across the topology, such as the number of flows competing at each link, the bottleneck link, and the fair share rate of each flow, etc. 
2) {\em temporal modeling}, which considers the evolution of the state over time, such as how remaining flow sizes and rates or queue lengths vary across time. 
Spatial and temporal dynamics are inter-related, but modeling them requires different forms of computation. 
Spatial modeling involves calculations on a graph (representing the network topology and flow routes), whereas temporal modeling requires localized state update equations (e.g., a queue's evolution as a function of its input rate and link capacity). 

Our key idea is to decompose the spatial and temporal dynamics of the model into two separate functions, each represented by a neural network with a suitable architecture for that task. 
This decomposition is inspired by the structure of existing flow-level simulators like \flowsim. 

To understand this structure, let us contrast \flowsim, summarized in \Cref{eq:flowsim}, to the general model in Equations~\eqref{eq:flow_level_model} and~\eqref{eq:general_model}. 
\flowsim makes a few simplifying assumptions.
First, \flowsim tracks only the remaining flow size \( s_f \) for each active flow \( f \) as the network state (no queue state). 
Second, it updates the remaining flow sizes using two key functions:
\begin{itemize}[leftmargin=*]
    \item \textbf{Spatial dynamics (\( f_{\text{space}} \)):} It computes the transmission rates \( R(t_i) = \big\{r_f \mid f \in \mathcal{F}(t_i)\big\} \) for all active flows based on the network topology and flow routes \( G(t_i) \). 
    This is done using a max-min fair rate allocation algorithm, which distributes bandwidth fairly among competing flows.
    
    \item \textbf{Temporal dynamics (\( f_{\text{time}} \)):} It updates the remaining size \( s_f(t_i) \) of each active flow by subtracting the amount of data transmitted since the last event. 
    The transmitted traffic is computed as the product of the previous transmission rate \( r_f(t_{i-1}) \) and the elapsed time \( t_i - t_{i-1} \).

\end{itemize}
Finally, \flowsim uses a query function \( g \) to estimate flow completion times (FCTs) \( O(t_i) \) by dividing the updated remaining size \( s_f(t_i) \) by the newly computed rate \( r_f(t_i) \) for each flow. 
The system of equations below summarizes \flowsim's operation as a discrete-time dynamical system:
\begin{equation}
\begin{aligned}
    X(t_i) &= \big\{ s_f(t_i) \mid f \in \mathcal{F}(t_i) \big\}, \\
    s_f(t_i) &= f_{\text{time}}\big( s_f(t_{i-1}), r_f(t_{i-1}), t_i - t_{i-1} \big), \forall f \in \mathcal{F}(t_i), \\
    R(t_i) &= f_{\text{space}}\big( X(t_i), G(t_i) \big), \\
    O(t_i) &= g\big( X(t_i), R(t_i) \big).
\end{aligned}
\label{eq:flowsim}
\end{equation}

\ours is a novel neural network architecture for modeling flow-level dynamics, inspired by \flowsim. 
It mimics the computational structure of \flowsim but replaces the hard-coded state update rules and calculations with learned components. 
\ours maintains a learnable ``hidden state'' for each network component, including individual flows and links. 
It does not specify the meaning of these hidden states; they are multidimensional vectors whose semantics are learned during the training process. 
\ours incorporates two key concepts based on \flowsim to define how the hidden states evolve during the simulation (\Cref{fig:motivation}):
\begin{enumerate}[leftmargin=*]
    \item \textbf{Spatial dynamics at each snapshot in time.}
    To model the spatial dynamics, \ours uses a Graph Neural Network (GNN) to model a function analogous to $f_{space}$ in Eq.~\eqref{eq:flowsim}. 
    Like $f_{space}$, \ours's GNN operates on a fixed graph capturing the interactions of active flows at one snapshot in time.  
    
     \item \textbf{Temporal dynamics on a per-component basis.}
     To model the temporal dynamics, \ours uses GRU~\cite{gru,gru-api} sequence models that evolve the hidden state of each component across time, modeling a function analogous to $f_{time}$ in Eq.~\eqref{eq:flowsim}. 
     Like $f_{time}$, \ours's GRU calculations do not require the graph information (topology and routes). 
     
\end{enumerate}

\vspace{1mm}
\noindent
\textbf{Benefits over \flowsim.} 
Although a model based on Eq.~\eqref{eq:flowsim} is more restrictive than the general model in Eq.~\eqref{eq:general_model}, \ours is still significantly more flexible than \flowsim. 
By replacing predefined states (e.g., remaining flow size) and spatial/temporal computations with multi-dimensional learnable hidden states and non-linear models, \ours can learn complex dynamics from the training data. 
For example, rather than assuming instantaneous max-min rate allocations, it can learn that the sending rate of a flow behaves differently during its startup and steady-state phases, e.g., by keeping track of the elapsed time since the start of each flow in its hidden state, and using this information in the GNN or GRU calculations. 
As our results will show, this flexibility allows \ours to capture the dynamics of congestion control and queuing much more accurately than \flowsim. 

\section{System Design of \ours}\label{sec-system-design}

\subsection{High-Level Overview}\label{subsec-system-design-1} 
\begin{figure}%[t!]
    \centering
    \includegraphics[width=0.99\linewidth]{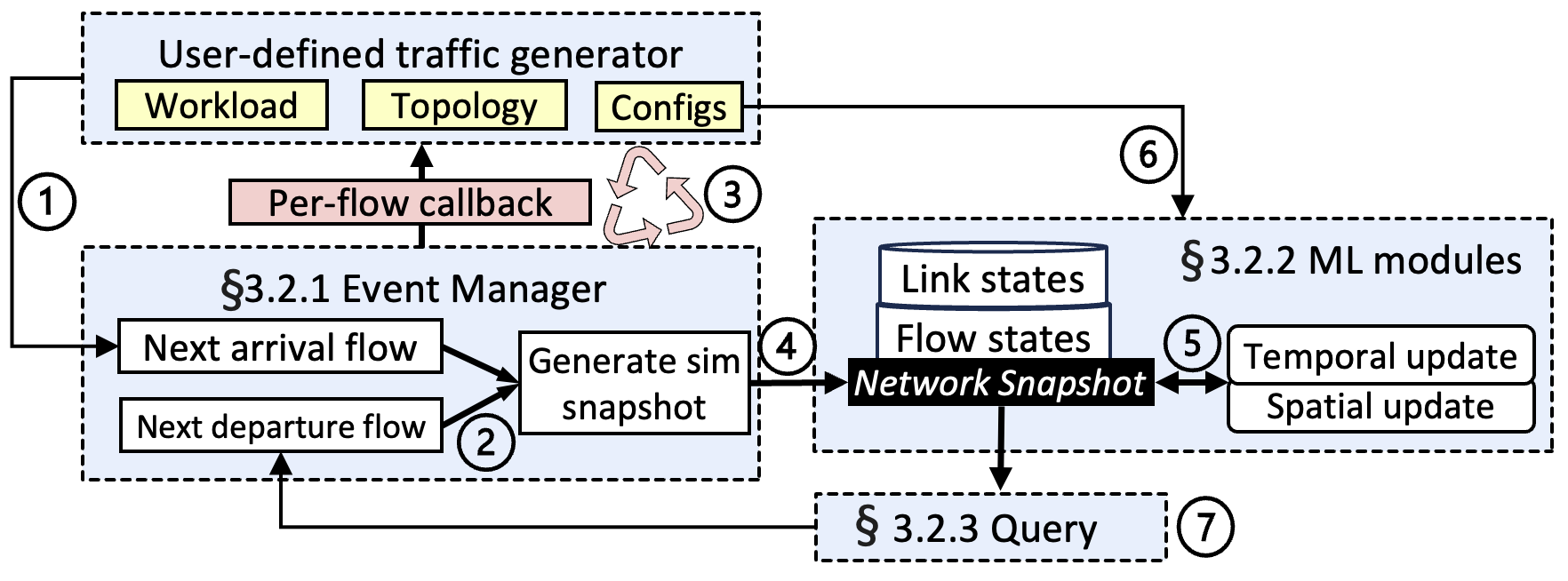}
    \caption{\small \ours’s workflow: Inputs (yellow boxes), outputs (red boxes), intermediate components (white boxes).}
    \vspace{-4mm}
    \label{fig:overview}
\end{figure}

Figure~\ref{fig:overview} illustrates \ours’s architecture.
\circled{1} Users provide a traffic workload and network topology to a traffic generator, which simulates different applications by continuously sending flows to \ours over time. 
Each flow includes its size, arrival time, and traversed network path.
A \emph{flow-level event} is either an arrival or departure -- both affect the dynamics of existing flows and therefore their completion times, so \ours must carefully track the order in which they occur.
\circled{2} At any given time, \ours can predict the completion times of all active flows, assuming no new flows arrive.
Based on the predicted next departure and the next arrival from the traffic generator, \ours’s event manager compares their times to determine which event occurs first. 
Then, it proceeds to the next event and updates its predictions for the departure times of active flows.
\ours processes the new event by generating a {\em network snapshot}. 
Rather than including the entire network state, a network snapshot only includes the subset of states affected by the new event, comprising any active flows and links impacted by the arrival or completion of a flow (\S\ref{subsubsec-system-design-2-1}).
\circled{3} For flow departures, \ours records the completion time and triggers callback functions to the traffic generator, which may initiate additional flows based on application logic.
\circled{4} \ours uses a ``hidden state'', a fixed-length learnable vector, to represent the state of each component (e.g., a single flow or an individual link) and maintains them throughout the simulation. 
\ours tracks flow-level dynamics by updating the hidden state of each component from \circled{5} two critical perspectives (\S\ref{subsubsec-system-design-2-2}).
(i) each component's \textbf{temporal update} since the last flow-level event (e.g., a flow’s traffic transmission),
and (ii) \textbf{spatial update} capturing interactions between flows and links at a snapshot in time, such as flows competing for bandwidth on shared links.
\circled{6} To account for diverse network configurations, like congestion control parameters and buffer sizes, \ours incorporates them as input parameters when updating flow states.
After simulating the snapshot, \circled{7} \ours queries the hidden states of all active flows and links to predict their behavior (\S\ref{subsubsec-system-design-2-3}). 
For example, it can estimate the remaining time-to-completion for a flow or the queue length for a specific link. 
The next flow departure event is sent to the event manager for processing.
This process iterates until the traffic generator completes sending all flows from the applications.

\subsection{\ours's Neural Network Architecture}\label{subsec-system-design-2}

\ours's ML architecture mimics the computational structure of \flowsim but replaces its hard-coded state update rules and calculations with learned components (\Cref{fig:motivation}).

\subsubsection{Handling network states}\label{subsubsec-system-design-2-1}
At any time $t_i$, \ours represents the network state \( X(t_i) \), which includes both flow and link states, as follows:
\begin{equation}
    X(t_i) = \big\{s_f \mid f \in \mathcal{F}(t_{i})\big\} \cup \big\{p_l \mid l \in \mathcal{L}(t_{i})\big\},
    \label{eq:state_m4}
\end{equation}
where \( s_f \) represents the state of each flow \( f \) in the active flow set \( \mathcal{F}(t_{i}) \), and \( p_l \) represents the state of each link \( l \) in the link set \( \mathcal{L}(t_{i}) \).

At the start of the simulation, \ours initializes the hidden states of all links based on their bandwidth. 
When a new flow arrives, \ours assigns it a hidden state, based on its flow size and the number of links it traverses. 
For every subsequent flow-level event, \ours dynamically updates the hidden states of both active flows and links. 

\subsubsection{Tracking network states}\label{subsubsec-system-design-2-2}
\ours tracks the network state of flows and links by updating their hidden state at each flow-level event. 
Whenever a new flow-level event occurs, it starts by generating a network snapshot, which only includes flows and links affected by the current event.
The update process involves two key steps:
First, \ours models the {\em temporal update} of each component using a sequence model (e.g., a single-layer LSTM or GRU). 
Specifically, \ours updates the hidden states of flows and links in the network snapshot, based on their previous hidden states and the elapsed time since the last event. 
This step is analogous to the remaining size updates in \flowsim (\( f_{\text{time}} \) in \Cref{eq:flowsim}).
Next, \ours models the {\em spatial interactions} between flows sharing network links using a graph neural network (GNN). 
To achieve this, \ours first converts the network snapshot into a bipartite graph where flow nodes represent active flows, and link nodes represent the network links. 
A bi-directional edge connects a flow node to a link node if the flow traverses that link.
\ours applies the GNN~\cite{hamilton2017inductive} to this bipartite graph to compute state updates that reflect the spatial interactions between components. 
This step mirrors the max-min fair rate calculations in \flowsim (\( f_{\text{space}} \) in \Cref{eq:flowsim}). 
We describe the bipartite graph construction and GNN operations in more detail in \S\ref{subsec-system-design-4}. 

\subsubsection{Querying flows and links}\label{subsubsec-system-design-2-3}
Finally, \ours predicts the FCT slowdowns of all active flows by querying their latest hidden state using a multilayer perceptron (MLP) model.
Specifically, \ours constructs a one-dimensional state vector by combining the latest hidden state of the flows, the number of links they traverse, and the network configuration parameters (e.g., congestion control algorithms and buffer sizes). 
For each active flow, the MLP takes its state vector as input and outputs the predicted flow completion time.
This step mirrors the simple completion time calculation in \flowsim (\( g \) in \Cref{eq:flowsim}).

\subsection{Adding Dense Supervision Signals}\label{subsec-system-design-3}
\begin{figure}%[t!]
    \centering
    \includegraphics[width=0.99\linewidth]{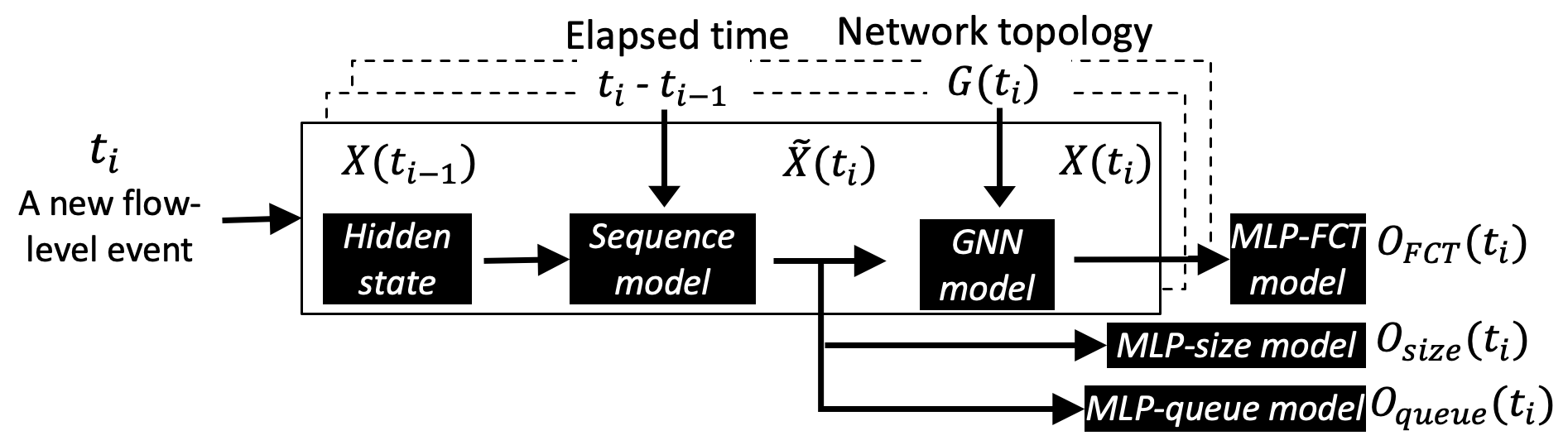}
    \caption{\small \ours adds ``dense'' supervision during training by querying intermediate network states for ``remaining size'' and ``queue length''. Dashed boxes represent subsequent simulations triggered by new flow-level events.}
    \vspace{-4mm}
    \label{fig:supervision}
\end{figure}

Training a complex model like \ours to generalize across diverse network scenarios requires learning the fundamental rules underlying flow-level network dynamics~\cite{dietmuller2022new}. 
To predict the FCT slowdowns of each flow, training \ours to minimize the mean L1 loss between predicted and actual FCT slowdowns for all flows provides a very sparse signal from which to learn. 
This challenge arises because \ours computes a sequence of updates to each flow's hidden state, one for each flow-level event during its lifetime. 
These sequences can be long; large flows may encounter hundreds or even thousands of flow-level events before completion.
Learning this entire process using only FCT information—essentially a single value per flow—is difficult. 
For large flows, a single prediction error must back-propagate through hundreds of intermediate state updates, further complicating training. 

Fortunately, \ours's ML architecture (\S\ref{subsec-system-design-2}) makes it straightforward to incorporate additional supervision signals to improve the quality of the learned functions.
Beyond predicting FCT, \ours can query intermediate states of flows and links for additional measurable attributes of the target system.
It can do this because it maintains hidden states for each flow and link which are meant to encompass all relevant information at any given time. 

\Cref{fig:supervision} shows how \ours uses the {\em remaining flow size} and {\em queue length} to provide ``dense'' supervision signals during training.
At each new flow-level event occurring at $t_{i}$, \ours retrieves the network states \( X(t_{i-1}) \) from the preceding event.
\ours first performs temporal updates using the sequence models, which take the previous hidden states \( X(t_{i-1}) \) and the elapsed time \( t_i - t_{i-1} \) as inputs. 
The updated hidden states \( \tilde X(t_i) \) represent the network's status immediately prior to the flow-level event at time $t_i$.
Next, \ours uses an MLP called \emph{MLP-size} to predict the remaining flow sizes based on the flow states \( \tilde X(t_i) \).
For each flow arrival event, \ours also queries the links traversed by this arriving flow to predict the queue length seen by its first packet, using another MLP model called \emph{MLP-queue}.
\ours predicts queue length only for the first packet of arriving flows for two reasons:
1) Queue length significantly influences the completion time of small flows (e.g., single-packet flows) but has a negligible impact on larger flows.
2) Limiting predictions to the first packet reduces training complexity and minimizes data collection overhead.

During training, \ours calculates three L1 losses for predicting FCT slowdowns, remaining sizes, and queue length and adds them into a single loss. 
\ours uses this combined signal to train the model.
By providing per-event supervision rather than relying solely on per-flow FCT, \ours learns more effectively and generalizes better across diverse scenarios.

\subsection{Using GNN on the Network Snapshot}\label{subsec-system-design-4}
In data center networks, flows compete for bandwidth on shared links, experience queueing delays, and are influenced by congestion control protocols and network configurations.
To capture these spatial interactions between flows and links, \ours incorporates a GNN into its ML architecture (\Cref{fig:overview}). 
The GNN updates the states of the affected flows and links at each flow-level event (\Cref{fig:supervision}).
\begin{equation}
\begin{aligned}
    X(t_i) & = f_{\text{space}}\big(\tilde X(t_i), G(t_i)\big), \\
\end{aligned}
\label{eq:m4}
\end{equation}
where \( f_{\text{space}} \) is modeled by a GNN in \ours, which performs multiple rounds of standard message passing operations~\cite{hamilton2017inductive} on a bipartite graph $G(t_i)$ constructed from the network snapshot. 
$G(t_i)$ includes only the flows affected by the flow-level event at \( t_i \) and the links they traverse, rather than all active flows.

\begin{figure}%[t!]
    \centering
    \includegraphics[width=0.8\linewidth]{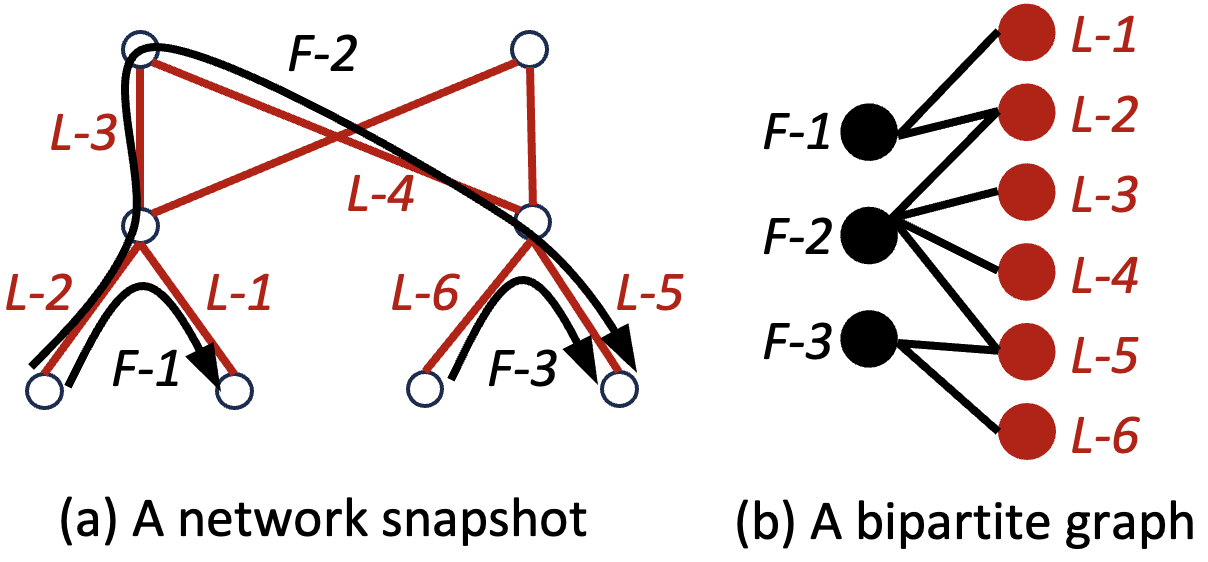}
    \vspace{-4mm}
    \caption{\small \ours converts (a) a network snapshot in time to a (b) bipartite graph and uses GNN to capture spatial dynamics.}
    \vspace{-4mm}
    \label{fig:gnn}
\end{figure}

\Cref{fig:gnn} shows how \ours converts a network snapshot to a bipartite graph.
Assuming that flow \( F_1 \) arrives, \ours considers a network snapshot that includes only flows \( F_1 \) to \( F_3 \), as these flows are directly or indirectly affected by \( F_1 \).
Other active flows in the network (not shown in the figure), which have no interactions with any flows affected by $F_1$, remain unchanged, as they are not impacted by the arrival of \( F_1 \).
To run the GNN, \ours converts this network snapshot into a bipartite graph consisting of three flow nodes (\( F_1 \) to \( F_3 \)) and six link nodes (\( L_1 \) to \( L_6 \)). 
A bidirectional edge is added between a flow and a link if the flow traverses that link. 
\ours initializes the features of both flow and link nodes using \( \tilde{X}(t_i) \) and performs several rounds of ``message passing'' GNN computations on the bipartite graph.   

The bipartite graph approach in \ours offers two key benefits for efficiency and generalization:
1) As the simulation progresses, \ours uses the GNN to only update the hidden states of the flows and links impacted by a given flow-level event, rather than recalculating the states for all active flows and links. 
This targeted update significantly improves the simulation efficiency. 
2) The GNN model in \ours is inherently topology-agnostic, operating on the ``flow-link'' bipartite graph. 
By training on smaller topologies that exhibit diverse flow-link interactions, the learned GNN generalizes effectively to larger and more complex topologies (\S\ref{sec-evaluation}).

To model the impact of various network configurations, \ours encodes network configuration into a one-dimensional vector representing key parameters, such as buffer size, initial window size, congestion control protocols (e.g., \dctcp~\cite{dctcp}, \timely~\cite{mittal2015timely}, \dcqcn~\cite{dcqcn}), and protocol parameters.
\ours provides this vector as an additional input to its underlying neural network models, including the GRU models and the MLP models used for prediction. 

\subsection{What \ours Does and Does Not Do}
\ours predicts individual flow behaviors, including FCT slowdown, remaining size over time, and queue length at switches traversed by the flow. 
These outputs also reflect overall network throughput and latency.
For instance, the FCT slowdown for short flows highlights packet latency and queuing delays, while the FCT of medium to long flows captures throughput effects. 

For speed, \ours operates at the flow level by assigning a static path to each flow for its entire lifetime. 
This design, however, restricts \ours from handling dynamic routing strategies like packet-spraying~\cite{spray} or flowlets~\cite{conga}.
Besides, \ours encodes congestion control protocols as one-hot vectors within the input features to the MLP model, limiting its ability to predict the performance of unseen CC protocols. 
Furthermore, the current implementation does not account for priority classes, which remains an area for future exploration.

\section{Implementation}\label{sec_implementation}
\vspace{1mm}
\noindent
\textbf{ML Models.}
\ours uses four GRUs~\cite{gru,gru-api} to update the 400-dimensional hidden states of flows and links in temporal updates.
For the spatial model, \ours uses a three-layer GraphSAGE GNN~\cite{gnn-api,hamilton2017inductive} with an embedding size of 300, using the ``sum'' aggregator.
\ours uses three two-layer MLPs with the hidden size of 200 to predict the per-flow FCT slowdown, remaining size, and the queue length. 
All models are trained from scratch using the PyTorch Lightning framework.
Training runs for 10 epochs on an A100 GPU with a batch size of 2 simulations. 
Each epoch takes about an hour. 
The size of model checkpoints is 6MBs. 
We use \nsthree version 3.39~\cite{ns3-3.39} to collect ground truth labels, which takes 1 to 2 minutes for every simulation in our training dataset (\S\ref{subsec-evaluation-1}).

\begin{figure}%[t!]
    \centering
    \includegraphics[width=0.99\linewidth]{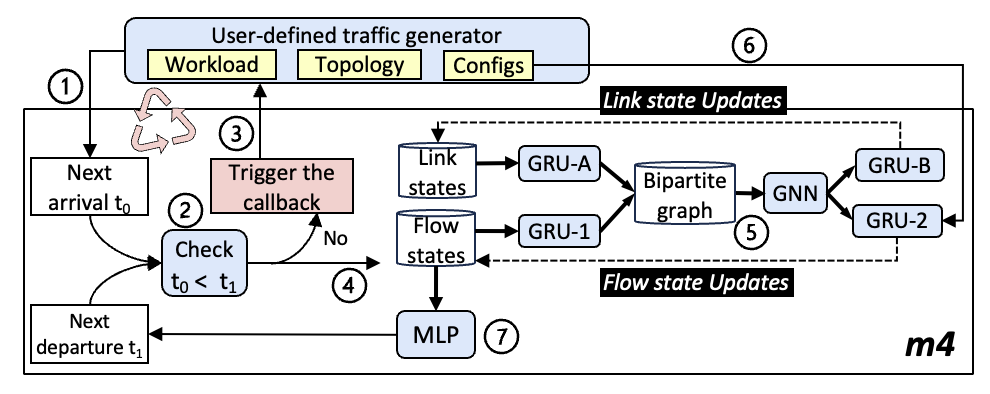}
    \vspace{-4mm}
    \caption{\small \ours's implementation}
    \vspace{-4mm}
    \label{fig:execution_pipeline}
\end{figure}

\vspace{1mm}
\noindent
\textbf{End-to-End Simulation.}
\ours implements its network simulation in C++, using LibTorch for ML inference.
We implemented a traffic generator in Rust as the application front end.
\Cref{fig:execution_pipeline} illustrates the interaction between the traffic generator and \ours as the network backend.
The simulation involves the following steps:
\circled{1} The traffic generator continuously sends flows to \ours for simulation.
\circled{2} \ours determines whether the next event is a flow arrival or departure based on their timestamp.
\circled{3} For a departure event, \ours triggers callback functions to the traffic generator and records the flow's completion time.
\circled{4} Using the current network states of flows and links, \ours first generates a network snapshot by selecting flows and links influenced by the current event and updates their hidden states using sequence models (GRU-A for links and GRU-1 for flows) for temporal updates since the last event.
Next, \circled{5} \ours converts this snapshot to a bipartite graph and runs GNN for spatial updates.  
\circled{6} \ours uses GRU-B and GRU-2 to further process the GNN's outputs alongside network configuration data, producing the final updated hidden states for both flows and links.
\circled{7} Finally, an MLP queries the updated hidden states of active flows to predict the next flow departure event.
This process repeats until all traffic is simulated. 
We run \ours on a single machine equipped with dual AMD EPYC 7763 64-core processors (256 CPUs and 512GB RAM) and an A100 GPU.

\section{Evaluation}\label{sec-evaluation}
We evaluate \ours's accuracy and scalability for large-scale network simulations (\S\ref{subsec-evaluation-2}) and its generalization across workloads and topologies (\S\ref{subsec-evaluation-3}). 
Then, we evaluate \ours's accuracy for predicting the performance of an interactive application as a closed-loop traffic generator in \S\ref{subsec-evaluation-4},
and dive deeper into the behavior of its dense supervision signals in \S\ref{subsec-evaluation-5}.

\begin{table}%[t!]
\centering
\begin{tabular}{l|l}
\toprule
\textbf{Parameter}            & \textbf{Sample space}                      \\
\hline
\small Synthetic flow size dist.     & \small Pareto, Exp, Gaussian, Log-normal            \\
Size parameter ($\theta$)     & \small $5K$ (small) to $50K$ (large), continuous \\
\hline
\small Empirical flow size dist.     & \small \cache, \web, \hadoop            \\
\hline
Burstiness param. ($\sigma$)         & $1$ (low) or $2$ (high) \\
Max link load             & $30$\% to $80$\%, continuous
\\ 
Traffic matrix & A, B, C~\cite{fb-network} \\
\small Oversubscription (plane) & 1-to-1, 2-to-1, 4-to-1 \\
\hline
{Init. window} & {5 to 15KB}, continuous \\
{Buffer size} & {100 to 160KB}, continuous\\
{CC protocol} & {DCTCP, TIMELY, DCQCN} \\
DCTCP ($K$) & 10 to 30KB, continuous\\
DCQCN ($K_{min}$, $K_{max}$) & (10 to 30KB, 30 to 50KB)\\
TIMELY ($T_{low}$, $T_{high}$) & (40 to 60$\mu$s, 100 to 150$\mu$s) \\
\bottomrule
\end{tabular}
\caption{\small The space of workloads and network configurations used in our dataset. 
The training set uses synthetic flow size distributions, while the test set uses empirical distributions.}
\vspace{-2mm}
\label{tab:parameters}
\end{table}

\subsection{Setup}\label{subsec-evaluation-1}
Each data point is created by randomly sampling workload and network configuration parameters from \Cref{tab:parameters}. 
We train \ours once and evaluate its performance in \S\ref{subsec-evaluation-2} to \S\ref{subsec-evaluation-5}.

\vspace{1mm}
\noindent
\textbf{Training Set} consists of 4,000 \nsthree simulations, each simulating 2,000 flows with flow size sampled from the synthetic distributions in \Cref{tab:parameters}.
The randomly sampled parameter $\theta$ adjusts the flow size scale for each simulation.
Inter-arrival times follow a log-normal distribution with two burstiness levels: $\sigma = 1$ for low burstiness and $\sigma = 2$ for high burstiness.
The max link load levels are chosen randomly to ensure that no link exceeds its capacity.
The traffic matrices represent rack-to-rack traffic, with hosts within each rack selected randomly for communication.
We use three cluster types~\cite{zhao2023scalable}: a database cluster (Matrix A), a web server cluster (Matrix~B), and a Hadoop cluster (Matrix C). 
We use an 8-rack, 32-host fat-tree topology.
All links have a capacity of 10Gbps, resulting in 4:1 oversubscription at the leaf level, similar to prior works~\cite{addanki2022abm,addanki2022powertcp,addanki2024credence,saeed2020annulus}. 
Each link has a propagation delay of 1$\mu$s.
Spine switches are organized into planes, and the oversubscription factor (1:1, 2:1, 4:1) is modulated at the plane level by varying the number of spines per plane.
We leave out 10\% of the simulations randomly for validation.

\vspace{1mm}
\noindent
\textbf{Real-world Test Set} uses the same setup as the training set except for the flow size distribution and the network topology.
For flow size distribution, we use the empirical distributions derived from Meta's data center network~\cite{fb-network} representing diverse applications such as databases (\cache), web servers (\web), and \hadoop. 
We use fat-tree network topologies at different scales.
In \S\ref{subsec-evaluation-2}, a large-scale 384-rack, 6144-host fat-tree topology, released by the study~\cite{fb-network} for Meta’s data center fabric design, is used to evaluate scalability. 
To verify \ours's scalability, we up-sample this topology up to a 49,152-host fat-tree.
Given the high computational complexity of running \nsthree to gather ground-truth data for the large setup, we scale down the topology and workload to a 64-rack, 1024-host fat-tree topology (\S\ref{subsec-evaluation-3}) and a 16-rack, 256-host fat-tree topology (\S\ref{subsec-evaluation-4}) for our sensitivity studies.

\vspace{1mm}
\noindent
\textbf{Baseline and Performance Metrics}. 
We compare \ours's performance with \flowsim, using \nsthree as the ground truth. 
The primary performance metric is the relative estimation error for per-flow FCT slowdown, defined as:
\begin{equation}
\label{eq:metric}
\frac{\text{estimated slowdown} - \text{ground-truth slowdown}}{\text{ground-truth slowdown}}
\end{equation}
The per-flow FCT slowdown serves as an indicator of \ours's accuracy.
For mean or \errortail of per-flow errors, we report the magnitude of the relative error, disregarding the sign.

\subsection{Large-scale Simulation}\label{subsec-evaluation-2}

\begin{table}
    \small
    \centering
    \begin{tabular}{c|c|c|c|c|c}
    \hline
    \multirow{ 2}{*}{Scenarios} & \multirow{ 2}{*}{Methods}  & \multicolumn{2}{c|}{\textbf{Per-flow error}} & \multirow{ 2}{*}{Time} & \multirow{ 2}{*}{Speedup}\\
     &  & mean & \errortail &  & \\
    \hline
    \multirow{3}{*}{\cache}
     & \nsthree & - & - & 3.6h & - \\
    & \flowsim & 22.4\% & 53.6\% & 59s & 220$\times$\\
    & \cellcolor{yellow!20} \ours & \cellcolor{yellow!20} 5.1\% & \cellcolor{yellow!20} 11.1\% & \cellcolor{yellow!20} 125s & \cellcolor{yellow!20} 104$\times$ \\
    \hline
    \multirow{3}{*}{\web}
     & \nsthree & - & - & 3.3h & - \\
    & \flowsim & 9.3\% & 39.9\%  & 57s & 208$\times$ \\
    & \cellcolor{yellow!20} \ours &  \cellcolor{yellow!20} 4.9\% & \cellcolor{yellow!20} 12.4\% & \cellcolor{yellow!20} 128s & \cellcolor{yellow!20} 92.8$\times$ \\
    \hline
    \multirow{3}{*}{\hadoop}
    & \nsthree & - & - & 3.5h & - \\
    & \flowsim & 12.4\% & 48.8\% & 57s & 221$\times$ \\
    & \cellcolor{yellow!20} \ours & \cellcolor{yellow!20} 3.7\% & \cellcolor{yellow!20} 9.3\% & \cellcolor{yellow!20} 129s & \cellcolor{yellow!20} 97.7$\times$ \\
    \hline
    \end{tabular}
    \caption{\small Comparison of \ours, \flowsim, and \nsthree for estimating per-flow FCT slowdown and runtime.}
    \label{tab:large_scale_sim}
    \vspace{-2mm}
\end{table}

\begin{table}
    \small
    \centering
    \begin{tabular}{c|c|c|c|c|c}
    \hline
   \# of hosts & 6,144 & 12,288 & 24,576 & 36,864 & 49,152 \\
   \hline
   \# of flows & 50,788 & 196,317 & 286,749 & 541,297 & 564,884 \\
    \hline
    \flowsim & 31s &54.2m &3.5h &23.5h & 29.5h  \\
    \hline
    \rowcolor{yellow!20}
    \ours & 138s &494s &701s &1530s &1413s \\
    Speedup & - & 6.58$\times$ & 18.0$\times$ & 55.3$\times$ & 75.2$\times$ \\
    GPU Mem. & 722 MiB & 1.8 GiB & 2.2 GiB & 4.5 GiB & 5.1 GiB \\
    \hline
    \end{tabular}
    \caption{\small Simulation runtime of \ours across large-scale network topologies with varying numbers of hosts.}
    \vspace{-2mm}
    \label{tab:runtime}
\end{table}

\vspace{1mm}
\noindent
\textbf{Setup.} We evaluate the scalability of \ours using a large-scale topology with 384 racks and 6,144 hosts~\cite{fb-network}. 
The topology comprises eight pods, each containing 48 racks and 16 hosts per rack, with a 2-to-1 oversubscription ratio at the plane level in the core network. 
The initial window size is 10KB, and the buffer size is set to 100KB. 
The simulations are conducted using \dctcp as the congestion control algorithm.

We evaluate three simulation scenarios with distinct empirical flow size distributions (\Cref{tab:large_scale_sim}), each with a maximum link load of 50\% and 50K flows per run.
We use the traffic matrix~B with low traffic burstiness (\( \sigma = 1 \)).

To evaluate \ours's scalability, we scale up Meta's data fabric by increasing the number of pods, expanding from 12,288 to 49,152 hosts in a fat-tree topology.
Each simulation runs for one second, producing 50K to 550K flows based on the topology size.

\vspace{1mm}
\noindent
\textbf{Quantitative Results}: Table~\ref{tab:large_scale_sim} summarizes the performance of \ours, \flowsim, and \nsthree in predicting per-flow slowdown, as well as simulation runtime.
\ours's mean (\errortail) per-flow error reduction compared to \flowsim ranges from 47\% (69\%) to 77\% (81\%). 
\ours significantly accelerates simulations, achieving up to \speedupovernsthree speedup over \nsthree and reducing simulation time from four hours to about two minutes.
\Cref{tab:runtime} further presents \ours's runtime for large-scale networks up to a 49,152 fat-tree topology.
\ours simulates 1 second of network activity in just 1413 seconds on a 49,152-host fat-tree.
It maintains a runtime proportional to the number of flows, leveraging GPU parallelism at each flow-level event with small to moderate memory consumption.
Note that \ours's speedups in \Cref{tab:runtime} are super-linearly increasing in the number of hosts.

\begin{figure}%[!t]
\centering
\subfigure[flow size $\in$ (0, 1KB{]}]{\includegraphics[width=0.46\linewidth]{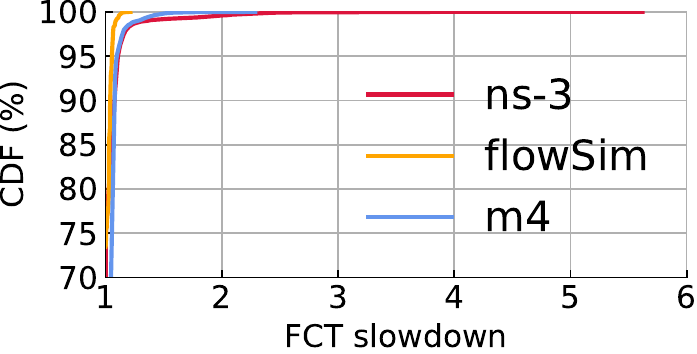}
\label{subfig:large_scale_error-1}}
\hspace{4mm}
% \hfill
\subfigure[flow size $\in$ (1KB, 200KB{]}]{\includegraphics[width=0.46\linewidth]{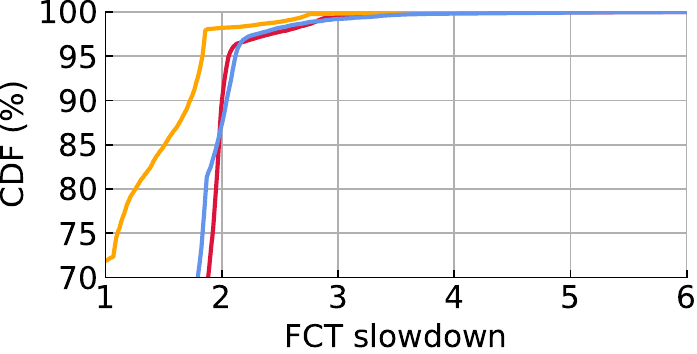}\label{subfig:large_scale_error-2}}
\subfigure[flow size $\in$ (200KB, 1MB{]}]{\includegraphics[width=0.46\linewidth]{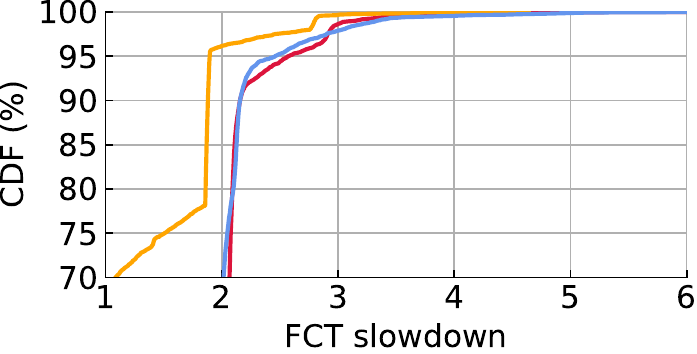}
\label{subfig:large_scale_error-3}}
\hspace{4mm}
% \hfill
\subfigure[flow size $\in$ (1MB, $\infty$)]{\includegraphics[width=0.46\linewidth]{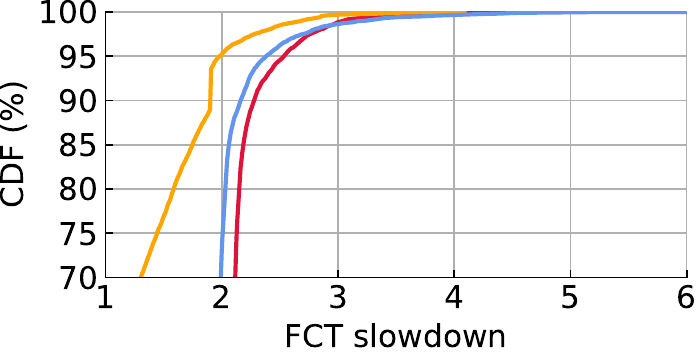}\label{subfig:large_scale_error-4}}
\vspace{-2mm}
\caption{\small CDF of per-flow FCT slowdown from \ours, \flowsim, and \nsthree across flow sizes in a large-scale simulation.
}
\vspace{-2mm}
\label{fig:large_scale_error}
\end{figure}

\begin{figure}%[!t]
\centering
\subfigure[FCT slowdown for \flowsim]{\includegraphics[width=0.46\linewidth]{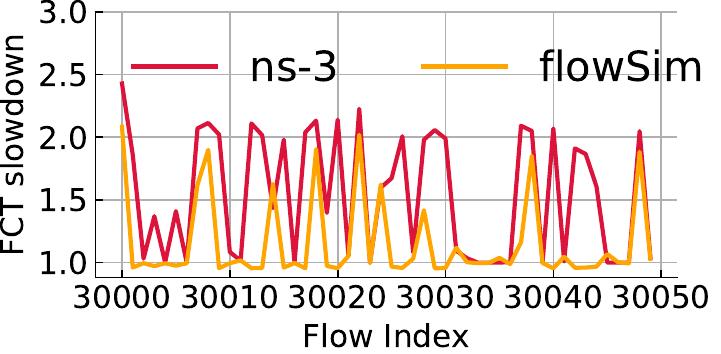}
\label{subfig:large_scale_fct-1}}
\hspace{4mm}
% \hfill
\subfigure[FCT slowdown for \ours]{\includegraphics[width=0.46\linewidth]{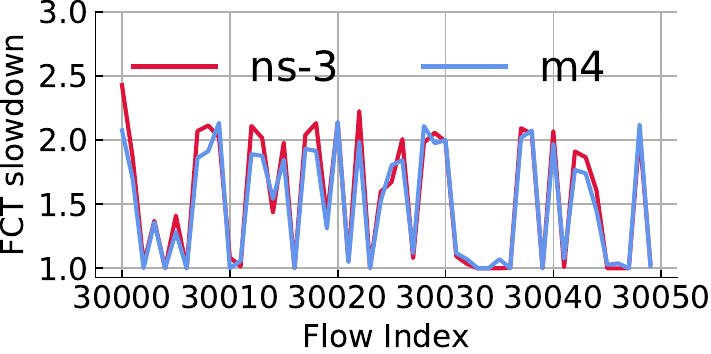}\label{subfig:large_scale_fct-2}}
\vspace{-2mm}
\caption{\small A sequence of FCT slowdowns estimated by \ours, \flowsim, and \nsthree in a large-scale network simulation.
}
\vspace{-2mm}
\label{fig:large_scale_fct}
\end{figure}

\vspace{1mm}
\noindent
\textbf{Comparative Insight}: \Cref{fig:large_scale_error} presents the CDF of per-flow FCT slowdown for \cache, comparing \ours, \flowsim, and \nsthree. 
\ours's estimations closely align with \nsthree across various flow size buckets, particularly for the tail slowdown. 
\Cref{fig:large_scale_fct} illustrates the predicted FCT slowdowns for 50 flows. 
\flowsim's predictions fail to capture the variations seen in \nsthree. 
In contrast, \ours accurately reflects the varying slowdowns, including the peak slowdown of flows 30,030 to 30,050. 

\begin{figure}%[!t]
\centering
\subfigure[Flow counts]{\includegraphics[width=0.32\linewidth]{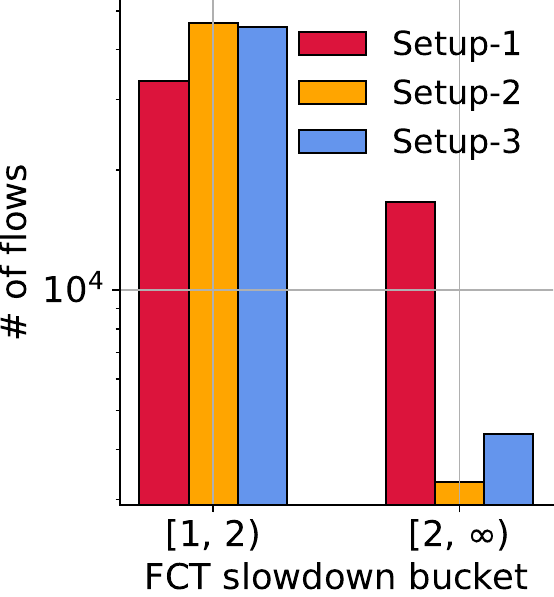}
\label{subfig:error_sldn-1}}
\hspace{-1mm}
\subfigure[\flowsim]{\includegraphics[width=0.32\linewidth]{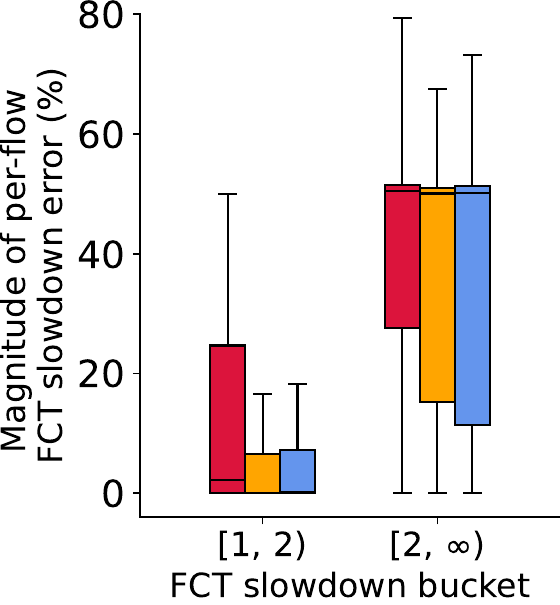}\label{subfig:error_sldn-2}}
\hspace{-1mm}
\subfigure[\ours]{\includegraphics[width=0.32\linewidth]{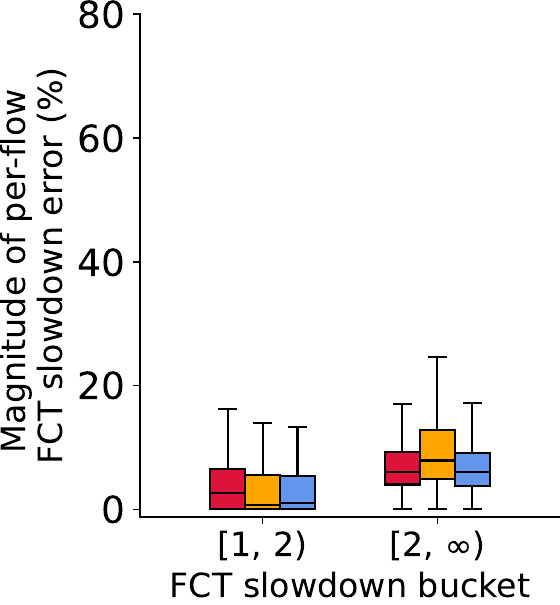}\label{subfig:error_sldn-3}}
\vspace{-2mm}
\caption{\small Error across slowdown buckets: \flowsim vs. \ours.}
\vspace{-2mm}
\label{fig:error_sldn}
\end{figure}

We further group the flows from each setup into various FCT slowdown buckets. 
\Cref{fig:error_sldn} presents the number of flows and the per-flow slowdown error distributions within each slowdown bucket.
\flowsim performs poorly for flows with large FCT slowdowns (\(>2\)), exhibiting a median error of approximately 40\% to 50\%.
In contrast, \ours significantly reduces the error for flows with slowdowns less than 2 to below 10\%. 
For flows with larger slowdowns, \ours achieves a median error of less than 20\%.

\subsection{Sensitivity Analysis}% at Small Scale}
\label{subsec-evaluation-3}

\noindent
\textbf{Setting.}
We evaluate \ours's robustness to various workloads, topologies, and network configurations using the small-scale 64-rack, 1024-host fat-tree topology described in \S\ref{subsec-evaluation-1}. 
This topology comprises four pods, each containing 16 racks and 16 hosts per rack, with variable spine counts to reflect different oversubscription levels at the plane level.
We randomly sample the parameter space from \Cref{tab:parameters} to generate 100 simulation scenarios.

\begin{figure}%[!t]
\centering
\subfigure[Mean error of FCT slowdown]{\includegraphics[width=0.46\linewidth]{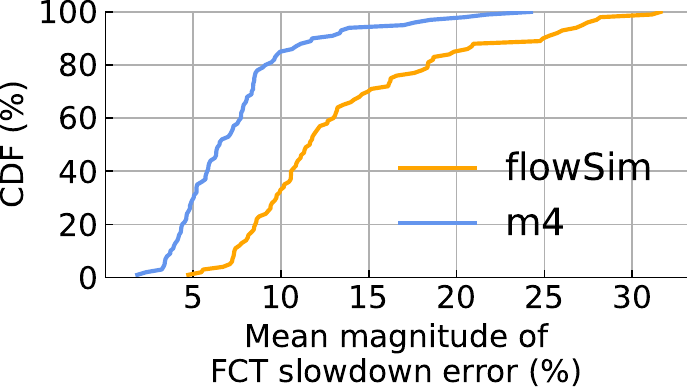}
\label{subfig:overall_perf-1}}
\hspace{4mm}
\subfigure[\errortail error of FCT slowdown]{\includegraphics[width=0.46\linewidth]{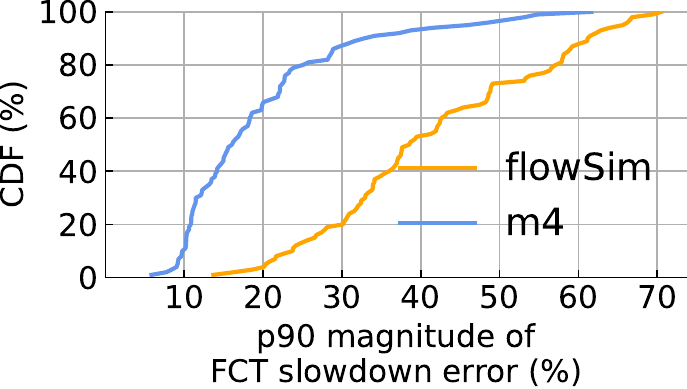}\label{subfig:overall_perf-2}}
\vspace{-2mm}
\caption{\small \ours is more accurate than \flowsim.
}
\vspace{-2mm}
\label{fig:overall_perf}
\end{figure}

\begin{figure}%[!t]
\centering
\subfigure[Traffic matrix]{\includegraphics[width=0.46\linewidth]{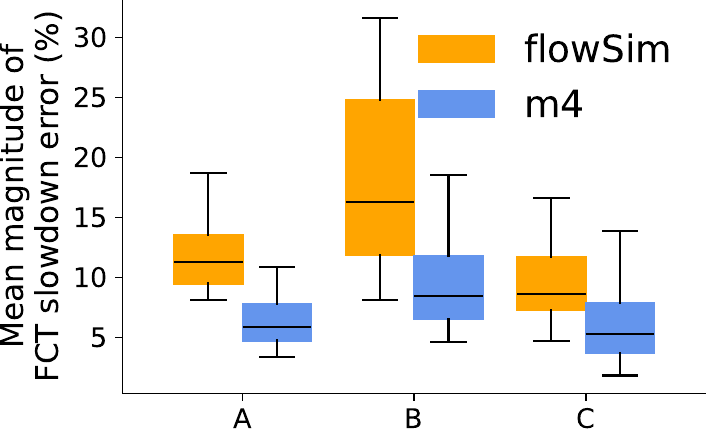}
\label{subfig:sensitivity_analysis-0}}
\hspace{4mm}
% \hfill
\subfigure[Flow size distribution]{\includegraphics[width=0.46\linewidth]{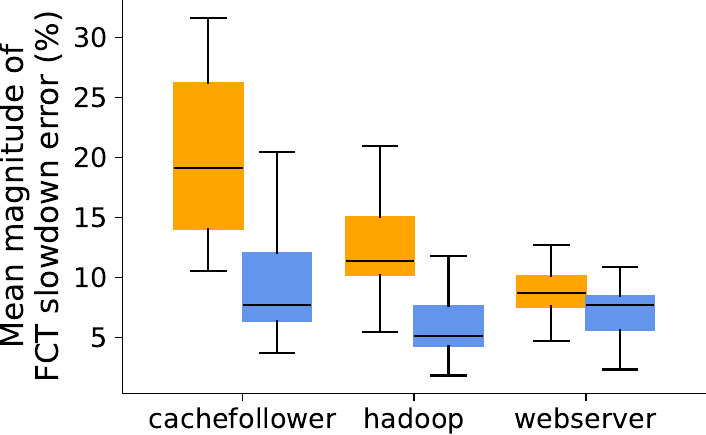}\label{subfig:sensitivity_analysis-1}}
% \vspace{-3mm}
\subfigure[Oversubscription]{\includegraphics[width=0.46\linewidth]{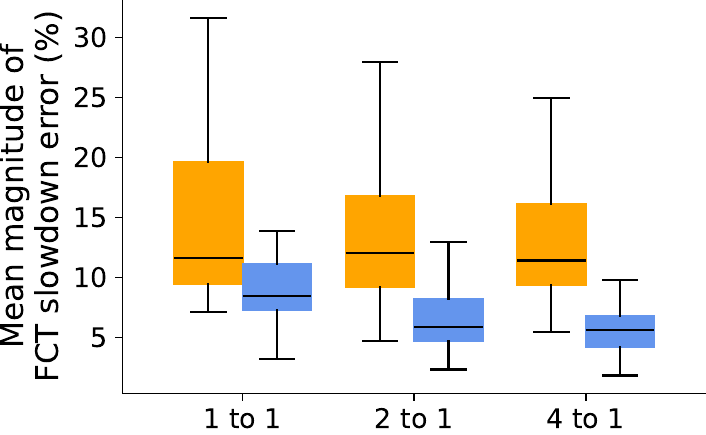}
\label{subfig:sensitivity_analysis-2}}
\hspace{4mm}
\subfigure[Max load (\%)]{\includegraphics[width=0.46\linewidth]{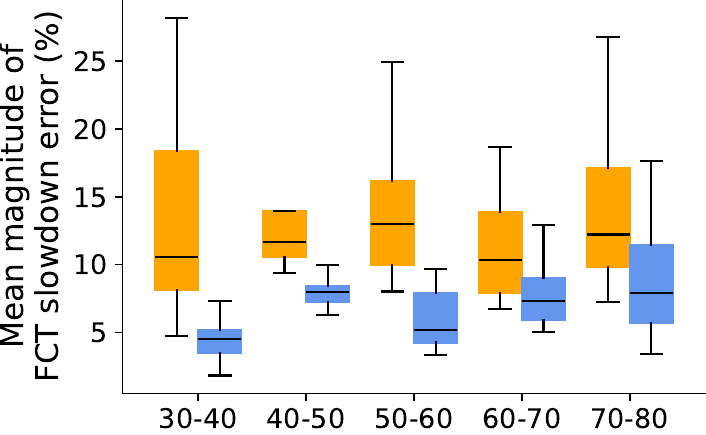}\label{subfig:sensitivity_analysis-3}}
% \vspace{-3mm}
\subfigure[Burstiness (log-normal's $\sigma$)]{\includegraphics[width=0.46\linewidth]{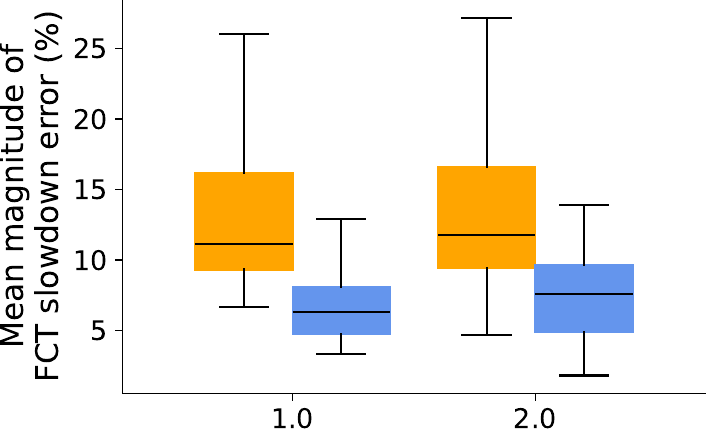}
\label{subfig:sensitivity_analysis-4}}
\hspace{4mm}
\subfigure[Congestion control algorithm]{\includegraphics[width=0.46\linewidth]{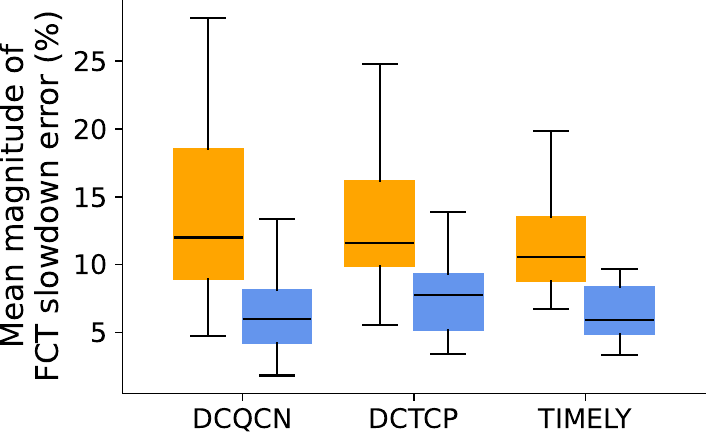}\label{subfig:sensitivity_analysis-5}}
% \vspace{-3mm}
\subfigure[Buffer size (KB)]{\includegraphics[width=0.46\linewidth]{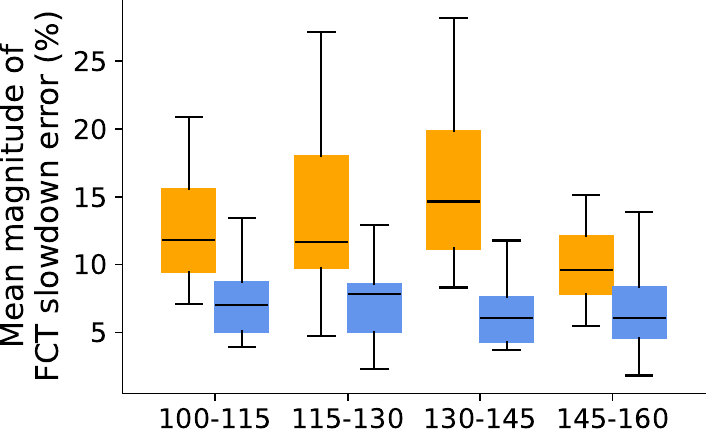}
\label{subfig:sensitivity_analysis-6}}
\hspace{4mm}
% \hfill
\subfigure[Init. window size (KB)]{\includegraphics[width=0.46\linewidth]{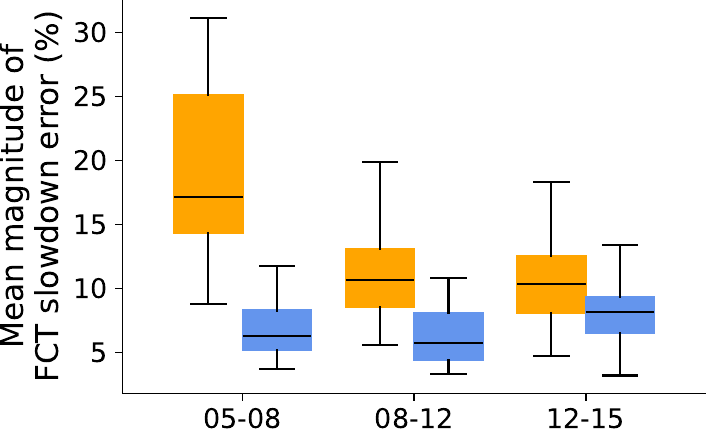}\label{subfig:sensitivity_analysis-7}}
\caption{\small Sensitivity of per-flow FCT slowdown error to parameters. Each boxplot depicts the distribution of the mean errors across 100 simulations for a configuration, with the center box capturing the middle 50\% (between the 25th and 75th percentiles) and a center line marking the median. Whiskers extend outwards to encompass the remaining data.
}
\vspace{-2mm}
\label{fig:sensitivity_analysis}
\end{figure}

\vspace{1mm}
\noindent
\textbf{Accuracy and Robustness.}
We calculate the mean and \errortail error of per-flow FCT slowdown for each scenario and present the CDF across 100 scenarios.
\Cref{subfig:overall_perf-1} shows the CDF of mean estimation errors for \ours and \flowsim. 
On average, \ours achieves a relative per-flow slowdown error of \errormeanours, significantly outperforming \flowsim's \errormeanflowsim. 
\Cref{subfig:overall_perf-2} highlights \ours's ability to predict challenging flows more accurately, achieving a \errortail error of \errortailours compared to \flowsim's \errortailflowsim.
\Cref{fig:sensitivity_analysis} illustrates the distribution of mean errors in per-flow slowdown estimation across different parameters.
\ours shows robust performance in estimating the FCT slowdown of each flow, regardless of variations in traffic matrix, flow size distribution, plane-level oversubscription, max load, burstiness, congestion control algorithm, buffer size, and initial congestion control window.
Both \flowsim and \ours experience slightly higher errors for the \cache flow size distribution, which has the largest flow sizes and leads to increased queue occupancy and queuing delays. 
However, \flowsim shows a more pronounced and skewed error pattern, particularly under traffic matrix B, a 1-to-1 oversubscription ratio, burstier workloads ($sigma$=2.0), and smaller initial window sizes (5–8 KB).

\subsection{Simulating an interactive application}\label{subsec-evaluation-4}

\noindent
\textbf{Setting.}
We evaluate \ours using a synthetic application with a closed-loop traffic generation pattern, where the number of inflight flows per rack is limited by a configurable parameter $N$.
This restriction introduces flow dependencies since new flows must wait for earlier flows to complete before starting.
This experiment uses a 16-rack, 256-host fat-tree topology (\S\ref{subsec-evaluation-1}). 
Hosts are divided into client and storage servers, with $1/4$ of the racks randomly assigned as client servers and the remainder as storage servers.
Size of flows between client and storage servers is generated randomly using the empirical workload distributions.
We assume clients have access to a backlog of flows, i.e., all flows are available to them at time zero.
We randomly generate 15 parameter configurations from \Cref{tab:parameters}, including various congestion control protocols and their associated settings. 
Each configuration is assigned with an inflight flow limit $N$, ranging from 1 to 13 in increments of 2, resulting in a total of 105 simulation scenarios.
For each scenario, we evaluate the overall throughput—measured as the number of completed flows per second—using \nsthree, \flowsim, and \ours.

\begin{figure}%[!t]
\centering
\subfigure[CDF of estimation error]{\includegraphics[width=0.46\linewidth]{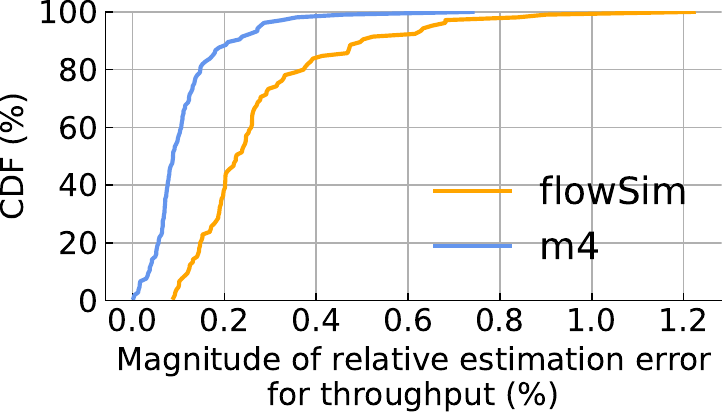}
\label{subfig:app-1}}
\hspace{4mm}
\subfigure[Throughput with various limits]{\includegraphics[width=0.46\linewidth]{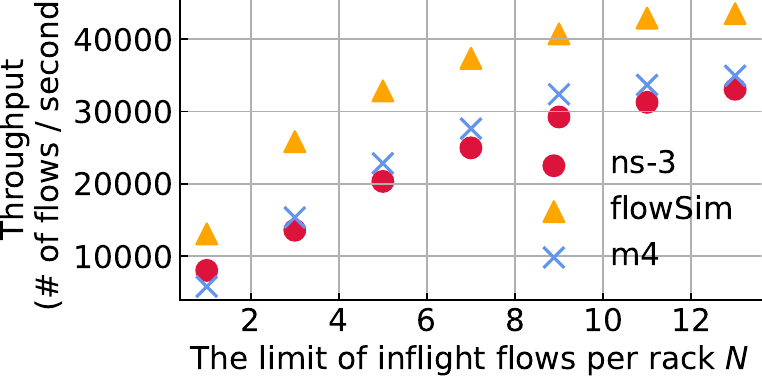}\label{subfig:app-2}}
\vspace{-2mm}
\caption{\small Simulation of an interactive application: (a) CDF of throughput estimation errors and (b) throughput vs flow limit per rack (N) for a specific scenario.}
\vspace{-2mm}
\label{fig:app}
\end{figure}

\vspace{1mm}
\noindent
\textbf{Accuracy and Robustness.} \Cref{subfig:app-1} compares the relative estimation error of throughput for \flowsim and \ours against \nsthree. 
\ours demonstrates significant improvements over \flowsim, reducing the mean (\errortail) throughput error from 28.1\% (49.9\%) to 11.5\% (22.3\%). 
\Cref{subfig:app-2} further illustrates the predicted throughput for a specific scenario across varying inflight flow limits \( N \).
\ours closely aligns with ns-3’s predictions across all limits.
However, \flowsim consistently overestimates throughput by underestimating the completion time of small flows due to ignoring congestion control and queuing effects.  

\vspace{1mm}
\noindent
\subsection{Dense Supervision}\label{subsec-evaluation-5}

\begin{figure}%[!t]
\centering
\subfigure[Mean error of FCT slowdown]{\includegraphics[width=0.46\linewidth]{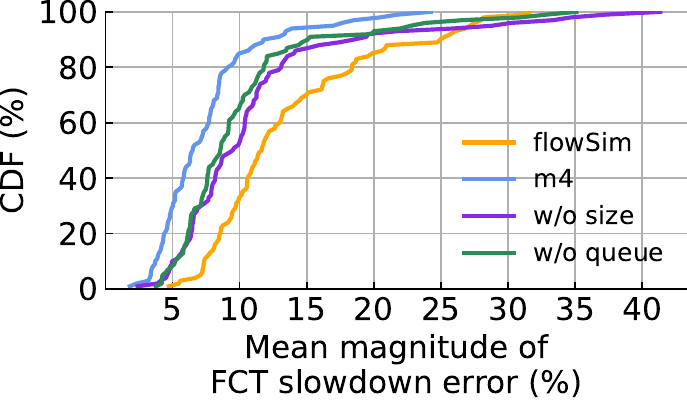}
\label{subfig:ablation_supervision-1}}
\hspace{4mm}
\subfigure[\errortail error of FCT slowdown]{\includegraphics[width=0.46\linewidth]{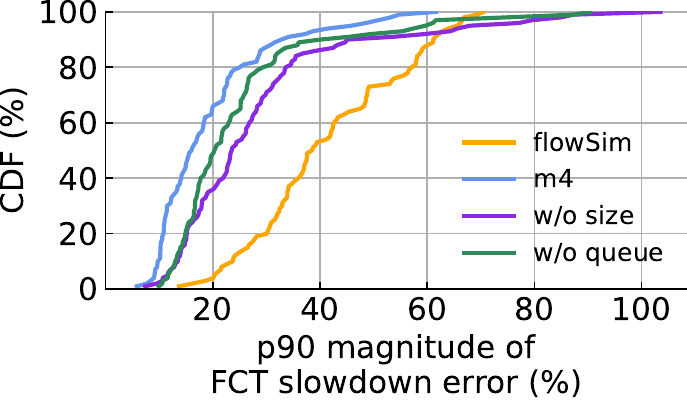}\label{subfig:ablation_supervision-2}}
\vspace{-2mm}
\caption{\small Ignoring dense supervision signals (remaining flow size or queue length) increases \ours's error.
}
\vspace{-2mm}
\label{fig:ablation_supervision}
\end{figure}

In this section, we dive deeper into the dense supervision signals in training \ours. 
\vspace{1mm}
\noindent
\textbf{Impact of removing dense supervision signals.} 
To assess the impact of dense supervision signals, such as predicting the remaining size and queue length, we exclude each signal during training \ours.
\Cref{tab:ablation_supervision} shows that removing the supervision signal for the remaining size increases the estimation error by 3.6\% (9.0\%) for the mean (\errortail) error and 6.2\% for the tail slowdown prediction. 
Similarly, excluding the supervision signal for queue length leads to a 2.6\% (5.5\%) increase in mean (tail) error and a 2.9\% increase in tail slowdown prediction.
The CDF of estimation errors, presented in \Cref{fig:ablation_supervision}, highlights that incorporating queue length prediction significantly improves tail FCT slowdown accuracy. 
This enhancement arises because the FCT of small flows is highly sensitive to queue length. 
Including queue length as a supervision signal enables faster learning of tail latency predictions, especially for small flows.

\begin{table}
    \centering
    \begin{tabular}{c|c|c|c}
    \hline
   Methods & Mean error& \errortail error& Tail sldn error \\
    \hline
    \flowsim  & \errormeanflowsim &\errortailflowsim & \errortailsldnflowsim \\
    \hline
    \rowcolor{yellow!20}
    \ours & \errormeanours &\errortailours & \errortailsldnours\\
    \hline
    w/o size & 11.1\% & 28.5\% & 18.5\% \\
    w/o queue & 10.1\% & 25.0\% & 16.2\% \\
    \hline
    % w/o \flowsim & 7.98\% & 11.18\% \\
    % \hline
    \end{tabular}
    \caption{\small Comparison of \ours, \flowsim, \ours without remaining size prediction and \ours without queue length prediction.}
    \vspace{-2mm}
    \label{tab:ablation_supervision}
\end{table}

\begin{figure}%[!t]
\centering
\subfigure[Error of remaining size]{\includegraphics[width=0.46\linewidth]{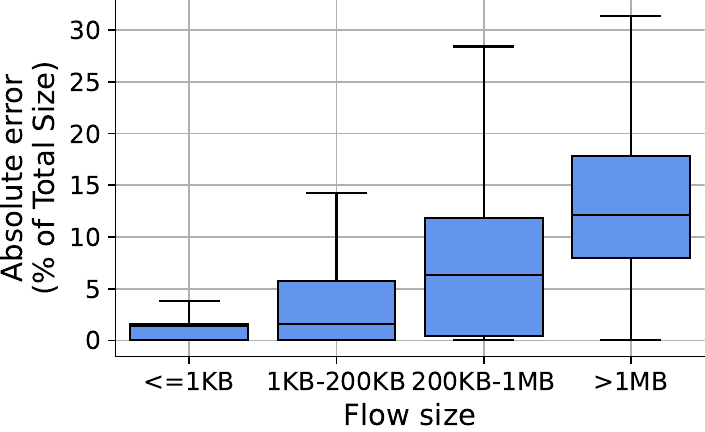}
\label{subfig:eva_supervision-1}}
\hspace{4mm}
% \hfill
\subfigure[Error of queue length]{\includegraphics[width=0.46\linewidth]{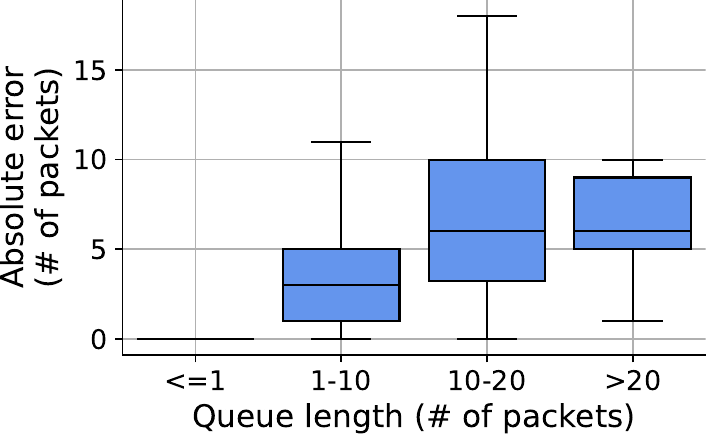}\label{subfig:eva_supervision-2}}
\subfigure[Remaining size prediction]{\includegraphics[width=0.46\linewidth]{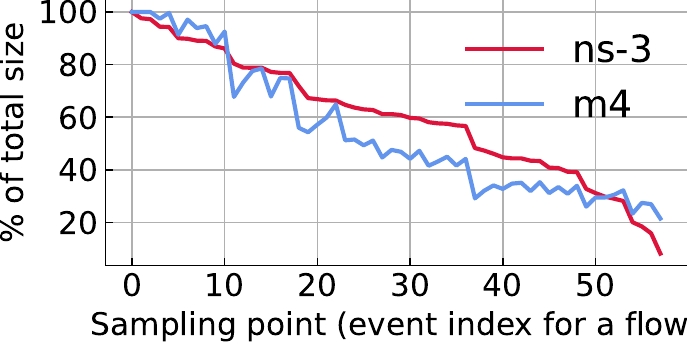}
\label{subfig:eva_supervision-3}}
\hspace{4mm}
\subfigure[Queue length prediction]{\includegraphics[width=0.46\linewidth]{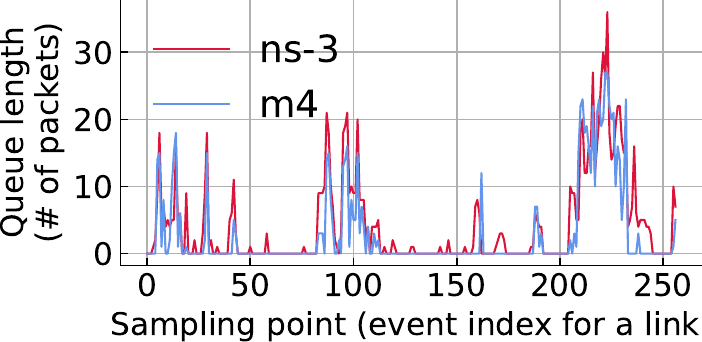}\label{subfig:eva_supervision-4}}
\vspace{-2mm}
\caption{\small \ours’s performance in predicting (a) remaining flow size and (b) queue length, compared to \nsthree. (c) and (d) provide examples of \ours’s predictions for a flow’s remaining size and a specific link’s queue length over time.
}
\vspace{-2mm}
\label{fig:eva_supervision}
\end{figure}

\vspace{1mm}
\noindent
\textbf{Predicting remaining size and queue length.}
\Cref{fig:eva_supervision} evaluates \ours's performance in predicting the remaining size of flows and the queue lengths.
For the remaining size prediction, \ours evaluates all active flows at each flow-level event across a test set of 100 scenarios. 
As shown in \Cref{subfig:eva_supervision-1}, \ours achieves an absolute error of less than 20\% for most flows (in terms of the total size of each flow). 
We think the error increases for larger flows due to the compounding error effect since they require frequent remaining size predictions due to their longer lifetimes and also interact with a large number of flows. 
\Cref{subfig:eva_supervision-3} illustrates an example of \ours's predictions for a single flow over 58 flow-level events. 
The predictions closely match the ground truth from \nsthree.
For queue length prediction, we focus on the queue length seen by the first packet of each flow during simulation. 
\Cref{subfig:eva_supervision-2} shows that \ours maintains a mean relative error of within 30\% across different queue lengths. 
Additionally, \Cref{subfig:eva_supervision-4} provides a trace of a specific link’s queue length over time. 
\ours effectively captures variations in queue length, closely aligning with the results from \nsthree.

\section{Related Work}\label{sec-related-work}

\textbf{Flow-level Simulation.} 
Flow-level simulation dynamically processes flow-level events to estimate individual flow behavior. 
\flowsim~\cite{massoulie1999bandwidth,namyar2023solving}, a standard flow-level model, efficiently estimates flow completion times by approximating congestion control as max-min bandwidth sharing and ignoring queue dynamics. 
It performs well for long flows but struggles with short flows, where queue dynamics significantly impact performance~\cite{li2024m3}.
\Citet{hockney1994communication,cao2024crux,rashidi2020astra} use the $\alpha$-$\beta$ model for network simulation, accounting for transmission delay as a combination of physical link delay and bandwidth-related delay. 
ASTRA-sim 2.0~\cite{won2023astra} extends this model by introducing FIFO queue scheduling for arriving flows. 
However, these methods lack support for various congestion control algorithms and bandwidth sharing across multiple jobs.

Several approaches use flow-level information to estimate aggregate performance but do not predict individual flow behavior. 
Fluid-based methods~\cite{misra2000,baccelli2003,marsan2004,DCTCPanalysis2011,peng2016} model flow evolution using partial differential equations to predict average behaviors in stochastic systems~\cite{eun2005}. 
Network Calculus~\cite{ciucu2012,le2001network} uses min-plus and max-plus algebras to provide worst-case bounds.
Machine learning methods such as Routenet~\cite{routenet2020,routeneterlang2022,routenetfermi2023} and m3~\cite{li2024m3} use GNNs or Transformers with flow-level inputs to predict aggregate performance.
Approaches like QT-Routenet~\cite{qtroutenet2022} extend Routenet by integrating predictions from queuing theory models as inputs to the GNN. 
RouteNet-Gauss~\cite{guemes2025routenet} is the most recent ML-based approach that learns flow-level dynamics from packet traces.
It models each type of component (flows, devices/queues, links) with an MLP and captures their interactions via a GNN. 
While fast, RouteNet requires a predefined trace and only predicts the distribution of flow-level performance within a fixed time window.
In contrast, \ours processes application flows on the fly and predicts the behavior of each flow.

\smallskip
\noindent
\textbf{Detailed Packet-level Simulation}\label{sub:related:low-level}
Packet-level simulators~\cite{ns3,opnet,omnet,htsim} model networks at the granularity of individual packets.
Their high computational demands make them impractical for large-scale data center networks. 
Parallelization techniques~\cite{paralleldes1990,parallelbook2001,nicol1994parallel,LIU2006genesis,synch2013} often yield limited speedups and, in some cases, even degrade performance~\cite{mimicnet,ns3}. 
DONS~\cite{dons} uses a data-oriented design to optimize multi-core, cache, and memory efficiency, achieving a 65$\times$ speedup on CPU-based clusters. 
Machine learning has been explored to accelerate packet-level simulations~\cite{kazer18fast}.
For example, DeepQueueNet~\cite{deepqueuenet} trains a model for each network component (links and switches) using packet-level simulations and can give packet-level telemetry as it operates on packets.
However, DeepQueueNet is still slow by operating at the packet level and achieves a 70$\times$ speedup on four V100 GPUs.
Besides, DeepQueueNet needs a pre-defined sequence of packets to operate on. 
Hence, it cannot respond to dynamic feedback, therefore being unable to model CCAs or closed-loop applications. 
Besides, MimicNet~\cite{mimicnet} learns network behavior from packet-level simulations by exploiting symmetries in fat-tree topologies~\cite{fattree2015}.
However, MimicNet does not handle abitrary topologies.
In contrast, \ours is a flow-level simulation that learns the behavior of CCAs without modeling packets and handles arbitrary topologies by using a bipartite graph modeling link and flow interactions. 
Furthermore, \ours processes flows as they are generated, thus handling closed-loop applications.
Network Traffic Transformer (NTT)~\cite{dietmuller2022new} uses Transformers to learn network dynamics from packet traces. 
While promising for simpler parking-lot topologies, NTT struggles with generalizing to diverse topologies and improving learning efficiency in varied environments. 
\ours tackles these challenges by proposing an ML architecture that effectively models network dynamics. \ours adds dense supervision signals to enhance learning.

Faster approaches~\cite{zhao2023scalable,song2024dex,song2024modelling} use parallelization to accelerate packet-level simulations for only aggregated network performance (e.g., tail latency, throughput).

\smallskip
\noindent
\textbf{Queuing Theory}\label{sub:related:high-level}
High-level simulations use queuing theory to model networks as systems of queues with stochastic arrival and service processes~\cite{book2000qtheory,bolch2006queueing}. 
While closed-form solutions exist under simplified assumptions like Poisson arrivals and single-packet flows, they fail to capture real-world complexities such as congestion control, bursty traffic, and diverse flow sizes.
MQL~\cite{mql,narayana2023similarity} refines queuing theory using regression trees to correct latency biases.
However, MQL assumes single-packet flows and relies on coarse inputs like average arrival rates, limiting its applicability to complex networks. 
More expressive models like the Markovian arrival process~\cite{map1993,chakravarthy2010markovian} improve accuracy but introduce computationally expensive state spaces, restricting them to performance estimation~\cite{Masuyama2003,KLEMM2003,ogras2010,kiasari2013,horvath2014commuting,mandal2020analytical,vranalysis2021map}.

\section{Conclusion}
\label{sec-discussion-conclusion}
We introduced \ours, a scalable and accurate ML-driven flow-level simulator that bridges the gap between flow- and packet-level simulations. 
\ours introduces a novel ML architecture that mimics the computational structure of conventional flow-level models while replacing hard-coded updates with learnable modules. 
To learn the underlying rules of network dynamics, \ours adds dense supervision by predicting intermediate network metrics such as remaining flow size and queue length during training. 
\ours achieves a speedup of up to \speedupovernsthree over packet-level simulation while reducing per-flow FCT slowdown estimation errors by \errormeanreduce (mean) and \errortailreduce (\errortail) compared to traditional flow-level models. 
Further, \ours generalizes across diverse data center network topologies, congestion control protocols, and workloads.

\section*{Acknowledgement}

This work was supported by NSF Career Award \#1751009, DARPA FastNICs program under contract \#HR0011-20-C-0089, and grants from Cisco Research and the UW Center for the Future of Cloud Infrastructure.

\bibliographystyle{ACM-Reference-Format}
\bibliography{reference}


\begin{thebibliography}{77}

%%% the Web address.

\ifx \showCODEN    \undefined \def \showCODEN     #1{\unskip}     \fi
\ifx \showDOI      \undefined \def \showDOI       #1{#1}\fi
\ifx \showISBNx    \undefined \def \showISBNx     #1{\unskip}     \fi
\ifx \showISBNxiii \undefined \def \showISBNxiii  #1{\unskip}     \fi
\ifx \showISSN     \undefined \def \showISSN      #1{\unskip}     \fi
\ifx \showLCCN     \undefined \def \showLCCN      #1{\unskip}     \fi
\ifx \shownote     \undefined \def \shownote      #1{#1}          \fi
\ifx \showarticletitle \undefined \def \showarticletitle #1{#1}   \fi
\ifx \showURL      \undefined \def \showURL       {\relax}        \fi
% invisible to TeX
\providecommand\bibfield[2]{#2}
\providecommand\bibinfo[2]{#2}
\providecommand\natexlab[1]{#1}
\providecommand\showeprint[2][]{arXiv:#2}

\bibitem[\protect\citeauthoryear{??}{ns3}{2024}]%
        {ns3-3.39}
 \bibinfo{year}{Retrieved by Dec 25th 2024}\natexlab{}.
\newblock \showarticletitle{ns-3.39}. In \bibinfo{booktitle}{{\em \url{https://www.nsnam.org/releases/ns-3-39/}}}.
\newblock


\bibitem[\protect\citeauthoryear{??}{grp}{2025}]%
        {grpc}
 \bibinfo{year}{Retrieved by Jan 28th 2025}\natexlab{}.
\newblock \showarticletitle{gRPC}. In \bibinfo{booktitle}{{\em \url{https://grpc.io/}}}.
\newblock


\bibitem[\protect\citeauthoryear{Addanki, Apostolaki, Ghobadi, Schmid, and Vanbever}{Addanki et~al\mbox{.}}{2022a}]%
        {addanki2022abm}
\bibfield{author}{\bibinfo{person}{Vamsi Addanki}, \bibinfo{person}{Maria Apostolaki}, \bibinfo{person}{Manya Ghobadi}, \bibinfo{person}{Stefan Schmid}, {and} \bibinfo{person}{Laurent Vanbever}.} \bibinfo{year}{2022}\natexlab{a}.
\newblock \showarticletitle{ABM: Active buffer management in datacenters}. In \bibinfo{booktitle}{{\em Proceedings of the ACM SIGCOMM}}. \bibinfo{pages}{36--52}.
\newblock


\bibitem[\protect\citeauthoryear{Addanki, Michel, and Schmid}{Addanki et~al\mbox{.}}{2022b}]%
        {addanki2022powertcp}
\bibfield{author}{\bibinfo{person}{Vamsi Addanki}, \bibinfo{person}{Oliver Michel}, {and} \bibinfo{person}{Stefan Schmid}.} \bibinfo{year}{2022}\natexlab{b}.
\newblock \showarticletitle{$\{$PowerTCP$\}$: Pushing the performance limits of datacenter networks}. In \bibinfo{booktitle}{{\em Proceedings of the USENIX NSDI}}. \bibinfo{pages}{51--70}.
\newblock


\bibitem[\protect\citeauthoryear{Addanki, Pacut, and Schmid}{Addanki et~al\mbox{.}}{2024}]%
        {addanki2024credence}
\bibfield{author}{\bibinfo{person}{Vamsi Addanki}, \bibinfo{person}{Maciej Pacut}, {and} \bibinfo{person}{Stefan Schmid}.} \bibinfo{year}{2024}\natexlab{}.
\newblock \showarticletitle{Credence: Augmenting Datacenter Switch Buffer Sharing with $\{$ML$\}$ Predictions}. In \bibinfo{booktitle}{{\em Proceedings of the USENIX NSDI}}. \bibinfo{pages}{613--634}.
\newblock


\bibitem[\protect\citeauthoryear{Alizadeh, Edsall, Dharmapurikar, Vaidyanathan, Chu, Fingerhut, Lam, Matus, Pan, Yadav, and Varghese}{Alizadeh et~al\mbox{.}}{2014}]%
        {conga}
\bibfield{author}{\bibinfo{person}{Mohammad Alizadeh}, \bibinfo{person}{Tom Edsall}, \bibinfo{person}{Sarang Dharmapurikar}, \bibinfo{person}{Ramanan Vaidyanathan}, \bibinfo{person}{Kevin Chu}, \bibinfo{person}{Andy Fingerhut}, \bibinfo{person}{Vinh~The Lam}, \bibinfo{person}{Francis Matus}, \bibinfo{person}{Rong Pan}, \bibinfo{person}{Navindra Yadav}, {and} \bibinfo{person}{George Varghese}.} \bibinfo{year}{2014}\natexlab{}.
\newblock \showarticletitle{CONGA: distributed congestion-aware load balancing for datacenters}. In \bibinfo{booktitle}{{\em Proceedings of the ACM SIGCOMM}}. \bibinfo{pages}{503–514}.
\newblock


\bibitem[\protect\citeauthoryear{Alizadeh, Greenberg, Maltz, Padhye, Patel, Prabhakar, Sengupta, and Sridharan}{Alizadeh et~al\mbox{.}}{2010}]%
        {dctcp}
\bibfield{author}{\bibinfo{person}{Mohammad Alizadeh}, \bibinfo{person}{Albert Greenberg}, \bibinfo{person}{David~A Maltz}, \bibinfo{person}{Jitendra Padhye}, \bibinfo{person}{Parveen Patel}, \bibinfo{person}{Balaji Prabhakar}, \bibinfo{person}{Sudipta Sengupta}, {and} \bibinfo{person}{Murari Sridharan}.} \bibinfo{year}{2010}\natexlab{}.
\newblock \showarticletitle{Data center tcp (dctcp)}. In \bibinfo{booktitle}{{\em Proceedings of the ACM SIGCOMM}}. \bibinfo{pages}{63–74}.
\newblock


\bibitem[\protect\citeauthoryear{Alizadeh, Javanmard, and Prabhakar}{Alizadeh et~al\mbox{.}}{2011}]%
        {DCTCPanalysis2011}
\bibfield{author}{\bibinfo{person}{Mohammad Alizadeh}, \bibinfo{person}{Adel Javanmard}, {and} \bibinfo{person}{Balaji Prabhakar}.} \bibinfo{year}{2011}\natexlab{}.
\newblock \showarticletitle{Analysis of DCTCP: stability, convergence, and fairness}. In \bibinfo{booktitle}{{\em Proceedings of the ACM SIGMETRICS Joint International Conference on Measurement and Modeling of Computer Systems}}. \bibinfo{pages}{73–84}.
\newblock


\bibitem[\protect\citeauthoryear{Alonso, Coll, Mart{\'\i}nez, Santonja, and L{\'o}pez}{Alonso et~al\mbox{.}}{2015}]%
        {fattree2015}
\bibfield{author}{\bibinfo{person}{Marina Alonso}, \bibinfo{person}{Salvador Coll}, \bibinfo{person}{Juan-Miguel Mart{\'\i}nez}, \bibinfo{person}{Vicente Santonja}, {and} \bibinfo{person}{Pedro L{\'o}pez}.} \bibinfo{year}{2015}\natexlab{}.
\newblock \showarticletitle{Power consumption management in fat-tree interconnection networks}.
\newblock \bibinfo{journal}{{\em Parallel computing\/}}  \bibinfo{volume}{48} (\bibinfo{year}{2015}), \bibinfo{pages}{59--80}.
\newblock


\bibitem[\protect\citeauthoryear{Andreyev}{Andreyev}{2025}]%
        {meta_fabric}
\bibfield{author}{\bibinfo{person}{Alexey Andreyev}.} \bibinfo{year}{Retrieved by Jan 28th 2025}\natexlab{}.
\newblock \showarticletitle{Introducing Data Center Fabric, the Next-Generation Facebook Data Center Network}. In \bibinfo{booktitle}{{\em \url{https://engineering.fb.com/2014/11/14/production-engineering/introducing-data-center-fabric-the-next-generation-facebook-data-center-network/}}}.
\newblock


\bibitem[\protect\citeauthoryear{Asmussen and Koole}{Asmussen and Koole}{1993}]%
        {map1993}
\bibfield{author}{\bibinfo{person}{S{\o}ren Asmussen} {and} \bibinfo{person}{Ger Koole}.} \bibinfo{year}{1993}\natexlab{}.
\newblock \showarticletitle{Marked point processes as limits of Markovian arrival streams}.
\newblock \bibinfo{journal}{{\em Journal of Applied Probability\/}} \bibinfo{volume}{30}, \bibinfo{number}{2} (\bibinfo{year}{1993}), \bibinfo{pages}{365--372}.
\newblock


\bibitem[\protect\citeauthoryear{Baccelli and Hong}{Baccelli and Hong}{2003}]%
        {baccelli2003}
\bibfield{author}{\bibinfo{person}{Fran{\c{c}}ois Baccelli} {and} \bibinfo{person}{Dohy Hong}.} \bibinfo{year}{2003}\natexlab{}.
\newblock \showarticletitle{Flow level simulation of large IP networks}. In \bibinfo{booktitle}{{\em Proceedings of the IEEE INFOCOM}}. \bibinfo{pages}{1911--1921}.
\newblock


\bibitem[\protect\citeauthoryear{Bolch, Greiner, De~Meer, and Trivedi}{Bolch et~al\mbox{.}}{2006}]%
        {bolch2006queueing}
\bibfield{author}{\bibinfo{person}{Gunter Bolch}, \bibinfo{person}{Stefan Greiner}, \bibinfo{person}{Hermann De~Meer}, {and} \bibinfo{person}{Kishor~S Trivedi}.} \bibinfo{year}{2006}\natexlab{}.
\newblock \bibinfo{booktitle}{{\em Queueing networks and Markov chains: modeling and performance evaluation with computer science applications}}.
\newblock \bibinfo{publisher}{John Wiley \& Sons, Ltd}. 821--868 pages.
\newblock
\showISBNx{9780471791577}
\showDOI{%
\url{https://doi.org/10.1002/0471791571.biblio}}


\bibitem[\protect\citeauthoryear{Cao, Guan, Qian, Gao, Xiao, Dong, Fu, Cai, and Zhai}{Cao et~al\mbox{.}}{2024}]%
        {cao2024crux}
\bibfield{author}{\bibinfo{person}{Jiamin Cao}, \bibinfo{person}{Yu Guan}, \bibinfo{person}{Kun Qian}, \bibinfo{person}{Jiaqi Gao}, \bibinfo{person}{Wencong Xiao}, \bibinfo{person}{Jianbo Dong}, \bibinfo{person}{Binzhang Fu}, \bibinfo{person}{Dennis Cai}, {and} \bibinfo{person}{Ennan Zhai}.} \bibinfo{year}{2024}\natexlab{}.
\newblock \showarticletitle{Crux: Gpu-efficient communication scheduling for deep learning training}. In \bibinfo{booktitle}{{\em Proceedings of the ACM SIGCOMM}}. \bibinfo{pages}{1--15}.
\newblock


\bibitem[\protect\citeauthoryear{Cao, Xia, Yang, Guo, Lu, Yuan, Zheng, Wu, Xiong, and Maltz}{Cao et~al\mbox{.}}{2013}]%
        {spray}
\bibfield{author}{\bibinfo{person}{Jiaxin Cao}, \bibinfo{person}{Rui Xia}, \bibinfo{person}{Pengkun Yang}, \bibinfo{person}{Chuanxiong Guo}, \bibinfo{person}{Guohan Lu}, \bibinfo{person}{Lihua Yuan}, \bibinfo{person}{Yixin Zheng}, \bibinfo{person}{Haitao Wu}, \bibinfo{person}{Yongqiang Xiong}, {and} \bibinfo{person}{Dave Maltz}.} \bibinfo{year}{2013}\natexlab{}.
\newblock \showarticletitle{Per-packet load-balanced, low-latency routing for clos-based data center networks}. In \bibinfo{booktitle}{{\em Proceedings of the ACM Conference on Emerging Networking Experiments and Technologies}}. \bibinfo{pages}{49–60}.
\newblock


\bibitem[\protect\citeauthoryear{Chakravarthy}{Chakravarthy}{2011}]%
        {chakravarthy2010markovian}
\bibfield{author}{\bibinfo{person}{Srinivas~R. Chakravarthy}.} \bibinfo{year}{2011}\natexlab{}.
\newblock \bibinfo{booktitle}{{\em Markovian Arrival Processes}}.
\newblock \bibinfo{publisher}{John Wiley \& Sons, Ltd}.
\newblock
\showISBNx{9780470400531}
\showDOI{%
\url{https://doi.org/10.1002/9780470400531.eorms0499}}


\bibitem[\protect\citeauthoryear{Cho, Kim, Choi, Heo, and Park}{Cho et~al\mbox{.}}{2024}]%
        {cho2024llmservingsim}
\bibfield{author}{\bibinfo{person}{Jaehong Cho}, \bibinfo{person}{Minsu Kim}, \bibinfo{person}{Hyunmin Choi}, \bibinfo{person}{Guseul Heo}, {and} \bibinfo{person}{Jongse Park}.} \bibinfo{year}{2024}\natexlab{}.
\newblock \showarticletitle{LLMServingSim: A HW/SW Co-Simulation Infrastructure for LLM Inference Serving at Scale}. In \bibinfo{booktitle}{{\em Proceedings of the IEEE International Symposium on Workload Characterization (IISWC)}}. \bibinfo{pages}{15--29}.
\newblock


\bibitem[\protect\citeauthoryear{Chung, Gulcehre, Cho, and Bengio}{Chung et~al\mbox{.}}{2014}]%
        {gru}
\bibfield{author}{\bibinfo{person}{Junyoung Chung}, \bibinfo{person}{Caglar Gulcehre}, \bibinfo{person}{KyungHyun Cho}, {and} \bibinfo{person}{Yoshua Bengio}.} \bibinfo{year}{2014}\natexlab{}.
\newblock \showarticletitle{Empirical evaluation of gated recurrent neural networks on sequence modeling}.
\newblock \bibinfo{journal}{{\em arXiv preprint arXiv:1412.3555\/}} (\bibinfo{year}{2014}).
\newblock


\bibitem[\protect\citeauthoryear{Ciucu and Schmitt}{Ciucu and Schmitt}{2012}]%
        {ciucu2012}
\bibfield{author}{\bibinfo{person}{Florin Ciucu} {and} \bibinfo{person}{Jens Schmitt}.} \bibinfo{year}{2012}\natexlab{}.
\newblock \showarticletitle{Perspectives on network calculus: no free lunch, but still good value}. In \bibinfo{booktitle}{{\em Proceedings of the ACM SIGCOMM}}. \bibinfo{pages}{311–322}.
\newblock


\bibitem[\protect\citeauthoryear{de~Aquino~Afonso and Berton}{de~Aquino~Afonso and Berton}{2022}]%
        {qtroutenet2022}
\bibfield{author}{\bibinfo{person}{Bruno~Klaus de Aquino~Afonso} {and} \bibinfo{person}{Lilian Berton}.} \bibinfo{year}{2022}\natexlab{}.
\newblock \showarticletitle{QT-Routenet: Improved GNN generalization to larger 5G networks by fine-tuning predictions from queueing theory}.
\newblock \bibinfo{journal}{{\em ITU Journal on Future and Evolving Technologies\/}} \bibinfo{volume}{3}, \bibinfo{number}{2} (\bibinfo{year}{2022}), \bibinfo{pages}{134–141}.
\newblock
\showISSN{2616-8375}
\showDOI{%
\url{https://doi.org/10.52953/fbrb3688}}


\bibitem[\protect\citeauthoryear{Dietm{\"u}ller, Ray, Jacob, and Vanbever}{Dietm{\"u}ller et~al\mbox{.}}{2022}]%
        {dietmuller2022new}
\bibfield{author}{\bibinfo{person}{Alexander Dietm{\"u}ller}, \bibinfo{person}{Siddhant Ray}, \bibinfo{person}{Romain Jacob}, {and} \bibinfo{person}{Laurent Vanbever}.} \bibinfo{year}{2022}\natexlab{}.
\newblock \showarticletitle{A new hope for network model generalization}. In \bibinfo{booktitle}{{\em Proceedings of the ACM HotNets}}. \bibinfo{pages}{152--159}.
\newblock


\bibitem[\protect\citeauthoryear{Eun}{Eun}{2005}]%
        {eun2005}
\bibfield{author}{\bibinfo{person}{Do~Young Eun}.} \bibinfo{year}{2005}\natexlab{}.
\newblock \showarticletitle{On the limitation of fluid-based approach for Internet congestion control}. In \bibinfo{booktitle}{{\em Proceedings of the International Conference On Computer Communications and Networks, {ICCCN}}}. \bibinfo{pages}{463--468}.
\newblock


\bibitem[\protect\citeauthoryear{Ferriol-Galm\'{e}s, Paillisse, Su\'{a}rez-Varela, Rusek, Xiao, Shi, Cheng, Barlet-Ros, and Cabellos-Aparicio}{Ferriol-Galm\'{e}s et~al\mbox{.}}{2023}]%
        {routenetfermi2023}
\bibfield{author}{\bibinfo{person}{Miquel Ferriol-Galm\'{e}s}, \bibinfo{person}{Jordi Paillisse}, \bibinfo{person}{Jos\'{e} Su\'{a}rez-Varela}, \bibinfo{person}{Krzysztof Rusek}, \bibinfo{person}{Shihan Xiao}, \bibinfo{person}{Xiang Shi}, \bibinfo{person}{Xiangle Cheng}, \bibinfo{person}{Pere Barlet-Ros}, {and} \bibinfo{person}{Albert Cabellos-Aparicio}.} \bibinfo{year}{2023}\natexlab{}.
\newblock \showarticletitle{RouteNet-Fermi: Network Modeling With Graph Neural Networks}.
\newblock \bibinfo{journal}{{\em IEEE/ACM Transactions on Networking\/}} \bibinfo{volume}{31}, \bibinfo{number}{6} (\bibinfo{year}{2023}), \bibinfo{pages}{3080–3095}.
\newblock
\showISSN{1063-6692}


\bibitem[\protect\citeauthoryear{Ferriol-Galm\'{e}s, Rusek, Su\'{a}rez-Varela, Xiao, Shi, Cheng, Wu, Barlet-Ros, and Cabellos-Aparicio}{Ferriol-Galm\'{e}s et~al\mbox{.}}{2022}]%
        {routeneterlang2022}
\bibfield{author}{\bibinfo{person}{Miquel Ferriol-Galm\'{e}s}, \bibinfo{person}{Krzysztof Rusek}, \bibinfo{person}{Jos\'{e} Su\'{a}rez-Varela}, \bibinfo{person}{Shihan Xiao}, \bibinfo{person}{Xiang Shi}, \bibinfo{person}{Xiangle Cheng}, \bibinfo{person}{Bo Wu}, \bibinfo{person}{Pere Barlet-Ros}, {and} \bibinfo{person}{Albert Cabellos-Aparicio}.} \bibinfo{year}{2022}\natexlab{}.
\newblock \showarticletitle{RouteNet-Erlang: A Graph Neural Network for Network Performance Evaluation}. In \bibinfo{booktitle}{{\em Proceedings of the IEEE INFOCOM}}. \bibinfo{pages}{2018--2027}.
\newblock


\bibitem[\protect\citeauthoryear{Fujimoto}{Fujimoto}{1990}]%
        {paralleldes1990}
\bibfield{author}{\bibinfo{person}{Richard~M Fujimoto}.} \bibinfo{year}{1990}\natexlab{}.
\newblock \showarticletitle{Parallel discrete event simulation}.
\newblock \bibinfo{journal}{{\it Commun. ACM}} \bibinfo{volume}{33}, \bibinfo{number}{10} (\bibinfo{year}{1990}), \bibinfo{pages}{30--53}.
\newblock


\bibitem[\protect\citeauthoryear{Fujimoto}{Fujimoto}{2001}]%
        {parallelbook2001}
\bibfield{author}{\bibinfo{person}{Richard~M Fujimoto}.} \bibinfo{year}{2001}\natexlab{}.
\newblock \showarticletitle{Parallel and distributed simulation systems}. In \bibinfo{booktitle}{{\em Proceeding of the Winter Simulation Conference}}. \bibinfo{pages}{147--157 vol.1}.
\newblock


\bibitem[\protect\citeauthoryear{Gao, Chen, Li, Liu, Wang, Zhang, and Lu}{Gao et~al\mbox{.}}{2023}]%
        {dons}
\bibfield{author}{\bibinfo{person}{Kaihui Gao}, \bibinfo{person}{Li Chen}, \bibinfo{person}{Dan Li}, \bibinfo{person}{Vincent Liu}, \bibinfo{person}{Xizheng Wang}, \bibinfo{person}{Ran Zhang}, {and} \bibinfo{person}{Lu Lu}.} \bibinfo{year}{2023}\natexlab{}.
\newblock \showarticletitle{DONS: Fast and Affordable Discrete Event Network Simulation with Automatic Parallelization}. In \bibinfo{booktitle}{{\em Proceedings of the ACM SIGCOMM}}. \bibinfo{pages}{167–181}.
\newblock


\bibitem[\protect\citeauthoryear{Geometric}{Geometric}{2024}]%
        {gnn-api}
\bibfield{author}{\bibinfo{person}{PyTorch Geometric}.} \bibinfo{year}{Retrieved by Dec 14th 2024}\natexlab{}.
\newblock \showarticletitle{conv.SAGEConv}. In \bibinfo{booktitle}{{\em \url{https://pytorch-geometric.readthedocs.io/en/stable/generated/torch_geometric.nn.conv.SAGEConv.html}}}.
\newblock


\bibitem[\protect\citeauthoryear{G{\"u}emes-Palau, Ferriol-Galm{\'e}s, Paillisse-Vilanova, L{\'o}pez-Bresc{\'o}, Barlet-Ros, and Cabellos-Aparicio}{G{\"u}emes-Palau et~al\mbox{.}}{2025}]%
        {guemes2025routenet}
\bibfield{author}{\bibinfo{person}{Carlos G{\"u}emes-Palau}, \bibinfo{person}{Miquel Ferriol-Galm{\'e}s}, \bibinfo{person}{Jordi Paillisse-Vilanova}, \bibinfo{person}{Albert L{\'o}pez-Bresc{\'o}}, \bibinfo{person}{Pere Barlet-Ros}, {and} \bibinfo{person}{Albert Cabellos-Aparicio}.} \bibinfo{year}{2025}\natexlab{}.
\newblock \showarticletitle{RouteNet-Gauss: Hardware-Enhanced Network Modeling with Machine Learning}.
\newblock \bibinfo{journal}{{\em arXiv preprint arXiv:2501.08848\/}} (\bibinfo{year}{2025}).
\newblock


\bibitem[\protect\citeauthoryear{Hamilton, Ying, and Leskovec}{Hamilton et~al\mbox{.}}{2017}]%
        {hamilton2017inductive}
\bibfield{author}{\bibinfo{person}{Will Hamilton}, \bibinfo{person}{Zhitao Ying}, {and} \bibinfo{person}{Jure Leskovec}.} \bibinfo{year}{2017}\natexlab{}.
\newblock \showarticletitle{Inductive representation learning on large graphs}.
\newblock \bibinfo{journal}{{\em Advances in neural information processing systems\/}}  \bibinfo{volume}{30} (\bibinfo{year}{2017}).
\newblock


\bibitem[\protect\citeauthoryear{Hochreiter}{Hochreiter}{1997}]%
        {lstm}
\bibfield{author}{\bibinfo{person}{S Hochreiter}.} \bibinfo{year}{1997}\natexlab{}.
\newblock \showarticletitle{Long Short-term Memory}.
\newblock \bibinfo{journal}{{\em Neural Computation MIT-Press\/}} (\bibinfo{year}{1997}).
\newblock


\bibitem[\protect\citeauthoryear{Hockney}{Hockney}{1994}]%
        {hockney1994communication}
\bibfield{author}{\bibinfo{person}{Roger~W Hockney}.} \bibinfo{year}{1994}\natexlab{}.
\newblock \showarticletitle{The communication challenge for MPP: Intel Paragon and Meiko CS-2}.
\newblock \bibinfo{journal}{{\em Parallel computing\/}} \bibinfo{volume}{20}, \bibinfo{number}{3} (\bibinfo{year}{1994}), \bibinfo{pages}{389--398}.
\newblock


\bibitem[\protect\citeauthoryear{Hong, Kandula, Mahajan, Zhang, Gill, Nanduri, and Wattenhofer}{Hong et~al\mbox{.}}{2013}]%
        {hong2013achieving}
\bibfield{author}{\bibinfo{person}{Chi-Yao Hong}, \bibinfo{person}{Srikanth Kandula}, \bibinfo{person}{Ratul Mahajan}, \bibinfo{person}{Ming Zhang}, \bibinfo{person}{Vijay Gill}, \bibinfo{person}{Mohan Nanduri}, {and} \bibinfo{person}{Roger Wattenhofer}.} \bibinfo{year}{2013}\natexlab{}.
\newblock \showarticletitle{Achieving high utilization with software-driven WAN}. In \bibinfo{booktitle}{{\em Proceedings of the ACM SIGCOMM}}. \bibinfo{pages}{15--26}.
\newblock


\bibitem[\protect\citeauthoryear{Horv{\'a}th, Van~Houdt, and Telek}{Horv{\'a}th et~al\mbox{.}}{2014}]%
        {horvath2014commuting}
\bibfield{author}{\bibinfo{person}{G{\'a}bor Horv{\'a}th}, \bibinfo{person}{B Van~Houdt}, {and} \bibinfo{person}{M Telek}.} \bibinfo{year}{2014}\natexlab{}.
\newblock \showarticletitle{Commuting matrices in the queue length and sojourn time analysis of MAP/MAP/1 queues}.
\newblock \bibinfo{journal}{{\em Stochastic Models\/}} \bibinfo{volume}{30}, \bibinfo{number}{4} (\bibinfo{year}{2014}), \bibinfo{pages}{554--575}.
\newblock


\bibitem[\protect\citeauthoryear{Inc}{Inc}{2024}]%
        {htsim}
\bibfield{author}{\bibinfo{person}{Broadcom Inc}.} \bibinfo{year}{Retrieved by Feb 1st 2024}\natexlab{}.
\newblock \showarticletitle{htsim Network Simulator}. In \bibinfo{booktitle}{{\em \url{https://github.com/Broadcom/csg-htsim}}}.
\newblock


\bibitem[\protect\citeauthoryear{Jafer, Liu, and Wainer}{Jafer et~al\mbox{.}}{2013}]%
        {synch2013}
\bibfield{author}{\bibinfo{person}{Shafagh Jafer}, \bibinfo{person}{Qi Liu}, {and} \bibinfo{person}{Gabriel Wainer}.} \bibinfo{year}{2013}\natexlab{}.
\newblock \showarticletitle{Synchronization methods in parallel and distributed discrete-event simulation}.
\newblock \bibinfo{journal}{{\em Simulation Modelling Practice and Theory\/}}  \bibinfo{volume}{30} (\bibinfo{year}{2013}), \bibinfo{pages}{54--73}.
\newblock
\showISSN{1569-190X}
\showDOI{%
\url{https://doi.org/10.1016/j.simpat.2012.08.003}}


\bibitem[\protect\citeauthoryear{Jain, Kumar, Mandal, Ong, Poutievski, Singh, Venkata, Wanderer, Zhou, Zhu, et~al\mbox{.}}{Jain et~al\mbox{.}}{2013}]%
        {jain2013b4}
\bibfield{author}{\bibinfo{person}{Sushant Jain}, \bibinfo{person}{Alok Kumar}, \bibinfo{person}{Subhasree Mandal}, \bibinfo{person}{Joon Ong}, \bibinfo{person}{Leon Poutievski}, \bibinfo{person}{Arjun Singh}, \bibinfo{person}{Subbaiah Venkata}, \bibinfo{person}{Jim Wanderer}, \bibinfo{person}{Junlan Zhou}, \bibinfo{person}{Min Zhu}, {et~al\mbox{.}}} \bibinfo{year}{2013}\natexlab{}.
\newblock \showarticletitle{B4: Experience with a globally-deployed software defined WAN}. In \bibinfo{booktitle}{{\em Proceedings of the ACM SIGCOMM}}. \bibinfo{pages}{3--14}.
\newblock


\bibitem[\protect\citeauthoryear{Kazer, Sedoc, Ng, Liu, and Ungar}{Kazer et~al\mbox{.}}{2018}]%
        {kazer18fast}
\bibfield{author}{\bibinfo{person}{Charles~W. Kazer}, \bibinfo{person}{Jo~{a}o Sedoc}, \bibinfo{person}{Kelvin~K.W. Ng}, \bibinfo{person}{Vincent Liu}, {and} \bibinfo{person}{Lyle~H. Ungar}.} \bibinfo{year}{2018}\natexlab{}.
\newblock \showarticletitle{Fast Network Simulation Through Approximation or: How Blind Men Can Describe Elephants}. In \bibinfo{booktitle}{{\em Proceedings of the ACM HotNets}}. \bibinfo{pages}{141–147}.
\newblock


\bibitem[\protect\citeauthoryear{Kiasari, Lu, and Jantsch}{Kiasari et~al\mbox{.}}{2013}]%
        {kiasari2013}
\bibfield{author}{\bibinfo{person}{Abbas~Eslami Kiasari}, \bibinfo{person}{Zhonghai Lu}, {and} \bibinfo{person}{Axel Jantsch}.} \bibinfo{year}{2013}\natexlab{}.
\newblock \showarticletitle{An Analytical Latency Model for Networks-on-Chip}.
\newblock \bibinfo{journal}{{\em IEEE Transactions on Very Large Scale Integration (VLSI) Systems\/}} \bibinfo{volume}{21}, \bibinfo{number}{1} (\bibinfo{year}{2013}), \bibinfo{pages}{113--123}.
\newblock
\showDOI{%
\url{https://doi.org/10.1109/TVLSI.2011.2178620}}


\bibitem[\protect\citeauthoryear{Klemm, Lindemann, and Lohmann}{Klemm et~al\mbox{.}}{2003}]%
        {KLEMM2003}
\bibfield{author}{\bibinfo{person}{Alexander Klemm}, \bibinfo{person}{Christoph Lindemann}, {and} \bibinfo{person}{Marco Lohmann}.} \bibinfo{year}{2003}\natexlab{}.
\newblock \showarticletitle{Modeling IP traffic using the batch Markovian arrival process}.
\newblock \bibinfo{journal}{{\em Performance Evaluation\/}} \bibinfo{volume}{54}, \bibinfo{number}{2} (\bibinfo{year}{2003}), \bibinfo{pages}{149--173}.
\newblock
\showISSN{0166-5316}
\showDOI{%
\url{https://doi.org/10.1016/S0166-5316(03)00067-1}}


\bibitem[\protect\citeauthoryear{Krishnaswamy, Singh, Mattes, Bissonnette, Bj{\o}rner, Nasrin, Kothari, Reddy, Abeln, Kandula, et~al\mbox{.}}{Krishnaswamy et~al\mbox{.}}{2023}]%
        {krishnaswamy2023onewan}
\bibfield{author}{\bibinfo{person}{Umesh Krishnaswamy}, \bibinfo{person}{Rachee Singh}, \bibinfo{person}{Paul Mattes}, \bibinfo{person}{Paul-Andre~C Bissonnette}, \bibinfo{person}{Nikolaj Bj{\o}rner}, \bibinfo{person}{Zahira Nasrin}, \bibinfo{person}{Sonal Kothari}, \bibinfo{person}{Prabhakar Reddy}, \bibinfo{person}{John Abeln}, \bibinfo{person}{Srikanth Kandula}, {et~al\mbox{.}}} \bibinfo{year}{2023}\natexlab{}.
\newblock \showarticletitle{$\{$OneWAN$\}$ is better than two: Unifying a split $\{$WAN$\}$ architecture}. In \bibinfo{booktitle}{{\em Proceedings of the USENIX NSDI}}. \bibinfo{pages}{515--529}.
\newblock


\bibitem[\protect\citeauthoryear{Le~Boudec and Thiran}{Le~Boudec and Thiran}{2001}]%
        {le2001network}
\bibfield{author}{\bibinfo{person}{Jean-Yves Le~Boudec} {and} \bibinfo{person}{Patrick Thiran}.} \bibinfo{year}{2001}\natexlab{}.
\newblock \bibinfo{booktitle}{{\em Network calculus: a theory of deterministic queuing systems for the internet}}.
\newblock \bibinfo{publisher}{Springer-Verlag}, \bibinfo{address}{Berlin, Heidelberg}.
\newblock
\showISBNx{354042184X}


\bibitem[\protect\citeauthoryear{Li, Nasr-Esfahany, Zhao, Noorbakhsh, Goyal, Alizadeh, and Anderson}{Li et~al\mbox{.}}{2024}]%
        {li2024m3}
\bibfield{author}{\bibinfo{person}{Chenning Li}, \bibinfo{person}{Arash Nasr-Esfahany}, \bibinfo{person}{Kevin Zhao}, \bibinfo{person}{Kimia Noorbakhsh}, \bibinfo{person}{Prateesh Goyal}, \bibinfo{person}{Mohammad Alizadeh}, {and} \bibinfo{person}{Thomas~E Anderson}.} \bibinfo{year}{2024}\natexlab{}.
\newblock \showarticletitle{m3: Accurate Flow-Level Performance Estimation using Machine Learning}. In \bibinfo{booktitle}{{\em Proceedings of the ACM SIGCOMM}}. \bibinfo{pages}{813--827}.
\newblock


\bibitem[\protect\citeauthoryear{Liu, Szymanski, and Saifee}{Liu et~al\mbox{.}}{2006}]%
        {LIU2006genesis}
\bibfield{author}{\bibinfo{person}{Yu Liu}, \bibinfo{person}{Boleslaw~K. Szymanski}, {and} \bibinfo{person}{Adnan Saifee}.} \bibinfo{year}{2006}\natexlab{}.
\newblock \showarticletitle{Genesis: A scalable distributed system for large-scale parallel network simulation}.
\newblock \bibinfo{journal}{{\em Computer Networks\/}} \bibinfo{volume}{50}, \bibinfo{number}{12} (\bibinfo{year}{2006}), \bibinfo{pages}{2028--2053}.
\newblock
\showISSN{1389-1286}
\showDOI{%
\url{https://doi.org/10.1016/j.comnet.2005.10.002}}


\bibitem[\protect\citeauthoryear{Lu and Yang}{Lu and Yang}{2012}]%
        {opnet}
\bibfield{author}{\bibinfo{person}{Zheng Lu} {and} \bibinfo{person}{Hongji Yang}.} \bibinfo{year}{2012}\natexlab{}.
\newblock \bibinfo{booktitle}{{\em Unlocking the power of OPNET modeler}}.
\newblock \bibinfo{publisher}{Cambridge University Press}. 1--238 pages.
\newblock
\showISBNx{9780521198745}
\showDOI{%
\url{https://doi.org/10.1017/CBO9780511667572}}


\bibitem[\protect\citeauthoryear{Mandal, Ayoub, Kishinevsky, Islam, and Ogras}{Mandal et~al\mbox{.}}{2021}]%
        {mandal2020analytical}
\bibfield{author}{\bibinfo{person}{Sumit~K. Mandal}, \bibinfo{person}{Raid Ayoub}, \bibinfo{person}{Micahel Kishinevsky}, \bibinfo{person}{Mohammad~M. Islam}, {and} \bibinfo{person}{Umit~Y. Ogras}.} \bibinfo{year}{2021}\natexlab{}.
\newblock \showarticletitle{Analytical Performance Modeling of NoCs under Priority Arbitration and Bursty Traffic}.
\newblock \bibinfo{journal}{{\em IEEE Embedded Systems Letters\/}} \bibinfo{volume}{13}, \bibinfo{number}{3} (\bibinfo{year}{2021}), \bibinfo{pages}{98--101}.
\newblock
\showDOI{%
\url{https://doi.org/10.1109/LES.2020.3013003}}


\bibitem[\protect\citeauthoryear{Marsan, Garetto, Giaccone, Leonardi, Schiattarella, and Tarello}{Marsan et~al\mbox{.}}{2004}]%
        {marsan2004}
\bibfield{author}{\bibinfo{person}{Marco~Ajmone Marsan}, \bibinfo{person}{Michele Garetto}, \bibinfo{person}{Paolo Giaccone}, \bibinfo{person}{Emilio Leonardi}, \bibinfo{person}{Enrico Schiattarella}, {and} \bibinfo{person}{Alessandro Tarello}.} \bibinfo{year}{2004}\natexlab{}.
\newblock \showarticletitle{Using partial differential equations to model TCP mice and elephants in large IP networks}. In \bibinfo{booktitle}{{\em Proceedings of the IEEE INFOCOM}}. \bibinfo{pages}{2821--2832 vol.4}.
\newblock


\bibitem[\protect\citeauthoryear{Massouli{\'e} and Roberts}{Massouli{\'e} and Roberts}{1999}]%
        {massoulie1999bandwidth}
\bibfield{author}{\bibinfo{person}{Laurent Massouli{\'e}} {and} \bibinfo{person}{James Roberts}.} \bibinfo{year}{1999}\natexlab{}.
\newblock \showarticletitle{Bandwidth sharing: objectives and algorithms}. In \bibinfo{booktitle}{{\em Proceedings of the IEEE INFOCOM}}. \bibinfo{pages}{1395--1403 vol.3}.
\newblock


\bibitem[\protect\citeauthoryear{Masuyama and Takine}{Masuyama and Takine}{2003}]%
        {Masuyama2003}
\bibfield{author}{\bibinfo{person}{Hiroyuki Masuyama} {and} \bibinfo{person}{Tetsuya Takine}.} \bibinfo{year}{2003}\natexlab{}.
\newblock \showarticletitle{Sojourn time distribution in a MAP/M/1 processor-sharing queue}.
\newblock \bibinfo{journal}{{\em Operations Research Letters\/}} \bibinfo{volume}{31}, \bibinfo{number}{5} (\bibinfo{year}{2003}), \bibinfo{pages}{406--412}.
\newblock
\showISSN{0167-6377}
\showDOI{%
\url{https://doi.org/10.1016/S0167-6377(03)00028-2}}


\bibitem[\protect\citeauthoryear{Misra, Gong, and Towsley}{Misra et~al\mbox{.}}{2000}]%
        {misra2000}
\bibfield{author}{\bibinfo{person}{Vishal Misra}, \bibinfo{person}{Wei-Bo Gong}, {and} \bibinfo{person}{Don Towsley}.} \bibinfo{year}{2000}\natexlab{}.
\newblock \showarticletitle{Fluid-based analysis of a network of AQM routers supporting TCP flows with an application to RED}. In \bibinfo{booktitle}{{\em Proceedings of the ACM SIGCOMM}}. \bibinfo{pages}{151–160}.
\newblock


\bibitem[\protect\citeauthoryear{Mittal, Lam, Dukkipati, Blem, Wassel, Ghobadi, Vahdat, Wang, Wetherall, and Zats}{Mittal et~al\mbox{.}}{2015}]%
        {mittal2015timely}
\bibfield{author}{\bibinfo{person}{Radhika Mittal}, \bibinfo{person}{Vinh~The Lam}, \bibinfo{person}{Nandita Dukkipati}, \bibinfo{person}{Emily Blem}, \bibinfo{person}{Hassan Wassel}, \bibinfo{person}{Monia Ghobadi}, \bibinfo{person}{Amin Vahdat}, \bibinfo{person}{Yaogong Wang}, \bibinfo{person}{David Wetherall}, {and} \bibinfo{person}{David Zats}.} \bibinfo{year}{2015}\natexlab{}.
\newblock \showarticletitle{TIMELY: RTT-based congestion control for the datacenter}.
\newblock \bibinfo{journal}{{\em Proceedings of the ACM SIGCOMM\/}}, \bibinfo{pages}{537–550}.
\newblock


\bibitem[\protect\citeauthoryear{Namyar, Arzani, Kandula, Segarra, Crankshaw, Krishnaswamy, Govindan, and Raj}{Namyar et~al\mbox{.}}{2024}]%
        {namyar2023solving}
\bibfield{author}{\bibinfo{person}{Pooria Namyar}, \bibinfo{person}{Behnaz Arzani}, \bibinfo{person}{Srikanth Kandula}, \bibinfo{person}{Santiago Segarra}, \bibinfo{person}{Daniel Crankshaw}, \bibinfo{person}{Umesh Krishnaswamy}, \bibinfo{person}{Ramesh Govindan}, {and} \bibinfo{person}{Himanshu Raj}.} \bibinfo{year}{2024}\natexlab{}.
\newblock \showarticletitle{Solving Max-Min Fair Resource Allocations Quickly on Large Graphs}. In \bibinfo{booktitle}{{\em Proceedings of the USENIX NSDI}}. \bibinfo{pages}{1937--1958}.
\newblock


\bibitem[\protect\citeauthoryear{Narayana, Shriver, O’Neal, Yildirim, Begaliyeva, and Ogras}{Narayana et~al\mbox{.}}{2023a}]%
        {narayana2023similarity}
\bibfield{author}{\bibinfo{person}{Shruti~Yadav Narayana}, \bibinfo{person}{Emily Shriver}, \bibinfo{person}{Kenneth O’Neal}, \bibinfo{person}{Nuriye Yildirim}, \bibinfo{person}{Khamida Begaliyeva}, {and} \bibinfo{person}{Umit~Y Ogras}.} \bibinfo{year}{2023}\natexlab{a}.
\newblock \showarticletitle{Similarity-Based Fast Analysis of Data Center Networks}.
\newblock \bibinfo{journal}{{\em IEEE Design \& Test\/}} (\bibinfo{year}{2023}).
\newblock


\bibitem[\protect\citeauthoryear{Narayana, Tong, Krishnakumar, Yildirim, Shriver, Ketkar, and Ogras}{Narayana et~al\mbox{.}}{2023b}]%
        {mql}
\bibfield{author}{\bibinfo{person}{Shruti~Yadav Narayana}, \bibinfo{person}{Jie Tong}, \bibinfo{person}{Anish Krishnakumar}, \bibinfo{person}{Nuriye Yildirim}, \bibinfo{person}{Emily Shriver}, \bibinfo{person}{Mahesh Ketkar}, {and} \bibinfo{person}{Umit~Y. Ogras}.} \bibinfo{year}{2023}\natexlab{b}.
\newblock \showarticletitle{MQL: ML-Assisted Queuing Latency Analysis for Data Center Networks}. In \bibinfo{booktitle}{{\em Proceedings of the IEEE International Symposium on Performance Analysis of Systems and Software (ISPASS)}}. \bibinfo{pages}{50--60}.
\newblock


\bibitem[\protect\citeauthoryear{Narayanan, Santhanam, Kazhamiaka, Phanishayee, and Zaharia}{Narayanan et~al\mbox{.}}{2020}]%
        {narayanan2020heterogeneity}
\bibfield{author}{\bibinfo{person}{Deepak Narayanan}, \bibinfo{person}{Keshav Santhanam}, \bibinfo{person}{Fiodar Kazhamiaka}, \bibinfo{person}{Amar Phanishayee}, {and} \bibinfo{person}{Matei Zaharia}.} \bibinfo{year}{2020}\natexlab{}.
\newblock \showarticletitle{$\{$Heterogeneity-Aware$\}$ cluster scheduling policies for deep learning workloads}. In \bibinfo{booktitle}{{\em Proceedings of the USENIX OSDI}}. \bibinfo{pages}{481--498}.
\newblock


\bibitem[\protect\citeauthoryear{Nicol and Fujimoto}{Nicol and Fujimoto}{1994}]%
        {nicol1994parallel}
\bibfield{author}{\bibinfo{person}{David Nicol} {and} \bibinfo{person}{Richard Fujimoto}.} \bibinfo{year}{1994}\natexlab{}.
\newblock \showarticletitle{Parallel simulation today}.
\newblock \bibinfo{journal}{{\em Annals of Operations Research\/}}  \bibinfo{volume}{53} (\bibinfo{year}{1994}), \bibinfo{pages}{249--285}.
\newblock


\bibitem[\protect\citeauthoryear{Ogras, Bogdan, and Marculescu}{Ogras et~al\mbox{.}}{2010}]%
        {ogras2010}
\bibfield{author}{\bibinfo{person}{Umit~Y. Ogras}, \bibinfo{person}{Paul Bogdan}, {and} \bibinfo{person}{Radu Marculescu}.} \bibinfo{year}{2010}\natexlab{}.
\newblock \showarticletitle{An Analytical Approach for Network-on-Chip Performance Analysis}.
\newblock \bibinfo{journal}{{\em IEEE Transactions on Computer-Aided Design of Integrated Circuits and Systems\/}} \bibinfo{volume}{29}, \bibinfo{number}{12} (\bibinfo{year}{2010}), \bibinfo{pages}{2001--2013}.
\newblock
\showDOI{%
\url{https://doi.org/10.1109/TCAD.2010.2061613}}


\bibitem[\protect\citeauthoryear{Peng, Walid, Hwang, and Low}{Peng et~al\mbox{.}}{2016}]%
        {peng2016}
\bibfield{author}{\bibinfo{person}{Qiuyu Peng}, \bibinfo{person}{Anwar Walid}, \bibinfo{person}{Jaehyun Hwang}, {and} \bibinfo{person}{Steven~H. Low}.} \bibinfo{year}{2016}\natexlab{}.
\newblock \showarticletitle{Multipath TCP: Analysis, Design, and Implementation}.
\newblock  \bibinfo{volume}{24}, \bibinfo{number}{1} (\bibinfo{year}{2016}), \bibinfo{pages}{596--609}.
\newblock
\showDOI{%
\url{https://doi.org/10.1109/TNET.2014.2379698}}


\bibitem[\protect\citeauthoryear{Peng, Zhang, Chen, and Zhang}{Peng et~al\mbox{.}}{2021}]%
        {vranalysis2021map}
\bibfield{author}{\bibinfo{person}{Xi Peng}, \bibinfo{person}{Fan Zhang}, \bibinfo{person}{Li Chen}, {and} \bibinfo{person}{Gong Zhang}.} \bibinfo{year}{2021}\natexlab{}.
\newblock \showarticletitle{A MAP-based Performance Analysis on 5G-powered Cloud VR Streaming}. In \bibinfo{booktitle}{{\em Proceedings of the IEEE International Conference on Communications (ICC)}}. \bibinfo{pages}{1--6}.
\newblock


\bibitem[\protect\citeauthoryear{PyTorch}{PyTorch}{2024}]%
        {gru-api}
\bibfield{author}{\bibinfo{person}{PyTorch}.} \bibinfo{year}{Retrieved by Dec 25th 2024}\natexlab{}.
\newblock \showarticletitle{nn.GRU}. In \bibinfo{booktitle}{{\em \url{https://pytorch.org/docs/stable/generated/torch.nn.GRU.html}}}.
\newblock


\bibitem[\protect\citeauthoryear{Rajasekaran, Ghobadi, and Akella}{Rajasekaran et~al\mbox{.}}{2024}]%
        {cassini}
\bibfield{author}{\bibinfo{person}{Sudarsanan Rajasekaran}, \bibinfo{person}{Manya Ghobadi}, {and} \bibinfo{person}{Aditya Akella}.} \bibinfo{year}{2024}\natexlab{}.
\newblock \showarticletitle{$\{$CASSINI$\}$:$\{$Network-Aware$\}$ Job Scheduling in Machine Learning Clusters}. In \bibinfo{booktitle}{{\em Proceedings of the USENIX NSDI}}. \bibinfo{pages}{1403--1420}.
\newblock


\bibitem[\protect\citeauthoryear{Rashidi, Sridharan, Srinivasan, and Krishna}{Rashidi et~al\mbox{.}}{2020}]%
        {rashidi2020astra}
\bibfield{author}{\bibinfo{person}{Saeed Rashidi}, \bibinfo{person}{Srinivas Sridharan}, \bibinfo{person}{Sudarshan Srinivasan}, {and} \bibinfo{person}{Tushar Krishna}.} \bibinfo{year}{2020}\natexlab{}.
\newblock \showarticletitle{Astra-sim: Enabling sw/hw co-design exploration for distributed dl training platforms}. In \bibinfo{booktitle}{{\em Proceedings of the IEEE International Symposium on Performance Analysis of Systems and Software (ISPASS)}}. \bibinfo{pages}{81--92}.
\newblock


\bibitem[\protect\citeauthoryear{Riley and Henderson}{Riley and Henderson}{2010}]%
        {ns3}
\bibfield{author}{\bibinfo{person}{George~F. Riley} {and} \bibinfo{person}{Thomas~R. Henderson}.} \bibinfo{year}{2010}\natexlab{}.
\newblock \bibinfo{booktitle}{{\em The ns-3 Network Simulator}}.
\newblock \bibinfo{publisher}{Springer Berlin Heidelberg}. 15--34 pages.
\newblock
\showISBNx{978-3-642-12331-3}
\showDOI{%
\url{https://doi.org/10.1007/978-3-642-12331-3_2}}


\bibitem[\protect\citeauthoryear{Robertazzi}{Robertazzi}{2000}]%
        {book2000qtheory}
\bibfield{author}{\bibinfo{person}{Thomas~G. Robertazzi}.} \bibinfo{year}{2000}\natexlab{}.
\newblock \bibinfo{booktitle}{{\em Computer Networks and Systems: Queueing Theory and Performance Evaluation}}.
\newblock \bibinfo{publisher}{Springer-Verlag}.
\newblock
\showISBNx{0387950370}


\bibitem[\protect\citeauthoryear{Roy, Zeng, Bagga, Porter, and Snoeren}{Roy et~al\mbox{.}}{2015}]%
        {fb-network}
\bibfield{author}{\bibinfo{person}{Arjun Roy}, \bibinfo{person}{Hongyi Zeng}, \bibinfo{person}{Jasmeet Bagga}, \bibinfo{person}{George Porter}, {and} \bibinfo{person}{Alex~C Snoeren}.} \bibinfo{year}{2015}\natexlab{}.
\newblock \showarticletitle{{Inside the Social Network's (Datacenter) Network}}. In \bibinfo{booktitle}{{\em Proceedings of the ACM SIGCOMM}}. \bibinfo{pages}{123–137}.
\newblock


\bibitem[\protect\citeauthoryear{Rusek, Su{\'a}rez-Varela, Almasan, Barlet-Ros, and Cabellos-Aparicio}{Rusek et~al\mbox{.}}{2020}]%
        {routenet2020}
\bibfield{author}{\bibinfo{person}{Krzysztof Rusek}, \bibinfo{person}{Jos{\'e} Su{\'a}rez-Varela}, \bibinfo{person}{Paul Almasan}, \bibinfo{person}{Pere Barlet-Ros}, {and} \bibinfo{person}{Albert Cabellos-Aparicio}.} \bibinfo{year}{2020}\natexlab{}.
\newblock \showarticletitle{RouteNet: Leveraging graph neural networks for network modeling and optimization in SDN}.
\newblock \bibinfo{journal}{{\em IEEE Journal on Selected Areas in Communications\/}} \bibinfo{volume}{38}, \bibinfo{number}{10} (\bibinfo{year}{2020}), \bibinfo{pages}{2260--2270}.
\newblock


\bibitem[\protect\citeauthoryear{Saeed, Gupta, Goyal, Sharif, Pan, Ammar, Zegura, Jang, Alizadeh, Kabbani, et~al\mbox{.}}{Saeed et~al\mbox{.}}{2020}]%
        {saeed2020annulus}
\bibfield{author}{\bibinfo{person}{Ahmed Saeed}, \bibinfo{person}{Varun Gupta}, \bibinfo{person}{Prateesh Goyal}, \bibinfo{person}{Milad Sharif}, \bibinfo{person}{Rong Pan}, \bibinfo{person}{Mostafa Ammar}, \bibinfo{person}{Ellen Zegura}, \bibinfo{person}{Keon Jang}, \bibinfo{person}{Mohammad Alizadeh}, \bibinfo{person}{Abdul Kabbani}, {et~al\mbox{.}}} \bibinfo{year}{2020}\natexlab{}.
\newblock \showarticletitle{Annulus: A dual congestion control loop for datacenter and wan traffic aggregates}. In \bibinfo{booktitle}{{\em Proceedings of the ACM SIGCOMM}}. \bibinfo{pages}{735--749}.
\newblock


\bibitem[\protect\citeauthoryear{Sharma, Bhasi, Singh, Kesidis, Kandemir, and Das}{Sharma et~al\mbox{.}}{2024}]%
        {sharma2024gpu}
\bibfield{author}{\bibinfo{person}{Aakash Sharma}, \bibinfo{person}{Vivek~M Bhasi}, \bibinfo{person}{Sonali Singh}, \bibinfo{person}{George Kesidis}, \bibinfo{person}{Mahmut~T Kandemir}, {and} \bibinfo{person}{Chita~R Das}.} \bibinfo{year}{2024}\natexlab{}.
\newblock \showarticletitle{GPU Cluster Scheduling for Network-Sensitive Deep Learning}.
\newblock \bibinfo{journal}{{\em arXiv preprint arXiv:2401.16492\/}} (\bibinfo{year}{2024}).
\newblock


\bibitem[\protect\citeauthoryear{Song, Garza, Meo, and Munaf{\`o}}{Song et~al\mbox{.}}{2024a}]%
        {song2024dex}
\bibfield{author}{\bibinfo{person}{Tailai Song}, \bibinfo{person}{Paolo Garza}, \bibinfo{person}{Michela Meo}, {and} \bibinfo{person}{Maurizio~M Munaf{\`o}}.} \bibinfo{year}{2024}\natexlab{a}.
\newblock \showarticletitle{DeX: Deep learning-based throughput prediction for real-time communications with emphasis on traffic eXtremes}.
\newblock \bibinfo{journal}{{\em Computer Networks\/}}  \bibinfo{volume}{249} (\bibinfo{year}{2024}), \bibinfo{pages}{110507}.
\newblock


\bibitem[\protect\citeauthoryear{Song, Garza, Meo, and Munaf{\`o}}{Song et~al\mbox{.}}{2024b}]%
        {song2024modelling}
\bibfield{author}{\bibinfo{person}{Tailai Song}, \bibinfo{person}{Paolo Garza}, \bibinfo{person}{Michela Meo}, {and} \bibinfo{person}{Maurizio~Matteo Munaf{\`o}}.} \bibinfo{year}{2024}\natexlab{b}.
\newblock \showarticletitle{Modelling Concurrent RTP Flows for End-to-end Predictions of QoS in Real Time Communications}.
\newblock \bibinfo{journal}{{\em arXiv preprint arXiv:2410.15846\/}} (\bibinfo{year}{2024}).
\newblock


\bibitem[\protect\citeauthoryear{Varga}{Varga}{2019}]%
        {omnet}
\bibfield{author}{\bibinfo{person}{Andras Varga}.} \bibinfo{year}{2019}\natexlab{}.
\newblock \bibinfo{booktitle}{{\em A Practical Introduction to the OMNeT++ Simulation Framework}}.
\newblock \bibinfo{publisher}{Springer International Publishing}. 3--51 pages.
\newblock
\showISBNx{978-3-030-12842-5}
\showDOI{%
\url{https://doi.org/10.1007/978-3-030-12842-5_1}}


\bibitem[\protect\citeauthoryear{Vaswani}{Vaswani}{2017}]%
        {transformer}
\bibfield{author}{\bibinfo{person}{A Vaswani}.} \bibinfo{year}{2017}\natexlab{}.
\newblock \showarticletitle{Attention is all you need}.
\newblock \bibinfo{journal}{{\em Advances in Neural Information Processing Systems\/}} (\bibinfo{year}{2017}).
\newblock


\bibitem[\protect\citeauthoryear{Won, Heo, Rashidi, Sridharan, Srinivasan, and Krishna}{Won et~al\mbox{.}}{2023}]%
        {won2023astra}
\bibfield{author}{\bibinfo{person}{William Won}, \bibinfo{person}{Taekyung Heo}, \bibinfo{person}{Saeed Rashidi}, \bibinfo{person}{Srinivas Sridharan}, \bibinfo{person}{Sudarshan Srinivasan}, {and} \bibinfo{person}{Tushar Krishna}.} \bibinfo{year}{2023}\natexlab{}.
\newblock \showarticletitle{Astra-sim2. 0: Modeling hierarchical networks and disaggregated systems for large-model training at scale}. In \bibinfo{booktitle}{{\em Proceedings of the IEEE International Symposium on Performance Analysis of Systems and Software (ISPASS)}}. \bibinfo{pages}{283--294}.
\newblock


\bibitem[\protect\citeauthoryear{Yang, Peng, Chen, Liu, Zhang, Xu, Li, and Zhang}{Yang et~al\mbox{.}}{2022}]%
        {deepqueuenet}
\bibfield{author}{\bibinfo{person}{Qingqing Yang}, \bibinfo{person}{Xi Peng}, \bibinfo{person}{Li Chen}, \bibinfo{person}{Libin Liu}, \bibinfo{person}{Jingze Zhang}, \bibinfo{person}{Hong Xu}, \bibinfo{person}{Baochun Li}, {and} \bibinfo{person}{Gong Zhang}.} \bibinfo{year}{2022}\natexlab{}.
\newblock \showarticletitle{DeepQueueNet: towards scalable and generalized network performance estimation with packet-level visibility}. In \bibinfo{booktitle}{{\em Proceedings of the ACM SIGCOMM}}. \bibinfo{pages}{441–457}.
\newblock


\bibitem[\protect\citeauthoryear{Zhang, Ng, Kazer, Yan, Sedoc, and Liu}{Zhang et~al\mbox{.}}{2021}]%
        {mimicnet}
\bibfield{author}{\bibinfo{person}{Qizhen Zhang}, \bibinfo{person}{Kelvin K.~W. Ng}, \bibinfo{person}{Charles Kazer}, \bibinfo{person}{Shen Yan}, \bibinfo{person}{Jo\~{a}o Sedoc}, {and} \bibinfo{person}{Vincent Liu}.} \bibinfo{year}{2021}\natexlab{}.
\newblock \showarticletitle{MimicNet: fast performance estimates for data center networks with machine learning}. In \bibinfo{booktitle}{{\em Proceedings of the ACM SIGCOMM}}. \bibinfo{pages}{287–304}.
\newblock


\bibitem[\protect\citeauthoryear{Zhao, Goyal, Alizadeh, and Anderson}{Zhao et~al\mbox{.}}{2023}]%
        {zhao2023scalable}
\bibfield{author}{\bibinfo{person}{Kevin Zhao}, \bibinfo{person}{Prateesh Goyal}, \bibinfo{person}{Mohammad Alizadeh}, {and} \bibinfo{person}{Thomas~E Anderson}.} \bibinfo{year}{2023}\natexlab{}.
\newblock \showarticletitle{Scalable Tail Latency Estimation for Data Center Networks}. In \bibinfo{booktitle}{{\em Proceedings of the USENIX NSDI}}. \bibinfo{pages}{685--702}.
\newblock


\bibitem[\protect\citeauthoryear{Zhu, Eran, Firestone, Guo, Lipshteyn, Liron, Padhye, Raindel, Yahia, and Zhang}{Zhu et~al\mbox{.}}{2015}]%
        {dcqcn}
\bibfield{author}{\bibinfo{person}{Yibo Zhu}, \bibinfo{person}{Haggai Eran}, \bibinfo{person}{Daniel Firestone}, \bibinfo{person}{ChaunXiong Guo}, \bibinfo{person}{Marina Lipshteyn}, \bibinfo{person}{Yehonatan Liron}, \bibinfo{person}{Jitendra Padhye}, \bibinfo{person}{Shachar Raindel}, \bibinfo{person}{Mohamad~Haj Yahia}, {and} \bibinfo{person}{Ming Zhang}.} \bibinfo{year}{2015}\natexlab{}.
\newblock \showarticletitle{Congestion Control for Large-Scale RDMA Deployments}. In \bibinfo{booktitle}{{\em Proceedings of the ACM SIGCOMM}}. \bibinfo{pages}{523–536}.
\newblock


\end{thebibliography}

\end{document}